\documentclass[12pt]{iopart} 

\usepackage{graphicx}
\usepackage{graphics}
\usepackage{amssymb}
\usepackage{epsfig}
\usepackage{latexsym}
\usepackage{color}
\usepackage{subfigure}

\newcommand{\bra}[1]{\left\langle{#1}\right\vert}
\newcommand{\ket}[1]{\left\vert{#1}\right\rangle}

\newcommand{\beq}{\begin{equation}}
\newcommand{\eeq}{\end{equation}}
\newcommand{\bqa}{\begin{eqnarray}}
\newcommand{\eqa}{\end{eqnarray}}
\newcommand{\nn}{\nonumber}

\newcommand{\erf}[1]{Eq.~(\ref{#1})}
\newcommand{\dg}{^\dagger}

\begin{document}

\title{Design principles and fundamental trade-offs in biomimetic light harvesting}
\author{Mohan Sarovar}
\address{Scalable and Secure Systems Research, Sandia National Laboratories, MS 9158, 7011 East Avenue, Livermore, California 94550 USA}
\ead{mnsarov@sandia.gov}
\author{K. Birgitta Whaley}
\address{Berkeley Center for Quantum Information and Computation, Berkeley, California 94720 USA}
\address{Department of Chemistry, University of California, Berkeley, California 94720 USA}


\begin{abstract}
Recent developments in synthetic and supramolecular chemistry have created opportunities to design organic systems with tailored nanoscale structure for various technological applications. A key application area is the capture of light energy and its conversion into electrochemical or chemical forms for photovoltaic or sensing applications. In this work we consider cylindrical assemblies of chromophores that model structures produced by several supramolecular techniques. Our study is especially guided by the versatile structures produced by virus-templated assembly. We use a multi-objective optimization framework to determine design principles and limitations in light harvesting performance for such assemblies, both in the presence and absence of disorder. We identify a fundamental trade-off in cylindrical assemblies that is encountered when attempting to maximize both efficiency of energy transfer and absorption bandwidth. We also rationalize the optimal design strategies and provide explanations for why various structures provide optimal performance. Most importantly, we find that the optimal design strategies depend on the amount of energetic and structural disorder in the system. The aim of these studies is to develop a program of quantum-informed rational design for construction of organic assemblies that have the same degree of tailored nanoscale structure as biological photosynthetic light harvesting complexes, and also have the potential to reproduce their remarkable light harvesting performance.
\end{abstract}

\maketitle

Interest in the molecular mechanisms underlying photosynthetic light harvesting has recently escalated, fueled in part by the potential of engineering \textit{biomimetic} solar energy harvesting technologies. The biomimetic approach aims to reproduce properties of the light harvesting complexes (LHCs) found in biology by using solid-state or organic components engineered at the nanoscale. Natural LHCs are remarkably efficient at all of the primary stages of photosynthesis: light capture, energy transfer, free carrier generation, and charge transfer \cite{Bla-2002}. Furthermore, LHCs perform these tasks in a manner that is robust to varying external and internal conditions. Reproducing such efficiencies and robustness would revolutionize our energy production capabilities. This constitutes part of the tremendous appeal of biomimetic approaches to designing photovoltaic technologies. However, photosynthetic light harvesting also has useful lessons for development of other technologies.  In particular, efficient photon capture and conversion to charge is also an important component of sensor technologies, raising the question whether these might also benefit from biomimetic approaches.

Biomimetic light harvesting may be viewed as one component, the ``front end", of artificial photosynthesis, which seeks to generate energy rich materials or fuels from sunlight.  Recent years have seen impressive achievements in several aspects of artificial photosynthesis, with progress in a diverse range of platforms ranging from molecular to semiconductor systems \cite{Anonymous:2006wf}. In this work we focus on light harvesting, the initial stage of any photosynthetic unit, and address the question of how the collection of light energy may be optimized by a biomimetic LHC and what tradeoffs are involved in achieving this goal, within a program of quantum-informed rational design.  

In order to construct a functional LHC from the bottom-up we require a detailed mapping between structural features and motifs, and the light harvesting or sensing capabilities of the composite system.  Establishing such a mapping is complicated by the wide variety of different structures evidenced by natural LHCS.  Despite the detailed variations in structure, however, some general motifs do emerge.  In particular, the majority of LHCs consist of densely packed pigment-protein complexes in which the light-absorbing chromophores (\textit{e.g.} chlorophyll and carotenoid molecules) are arranged with specific relative orientations and locations within scaffolds provided by proteins \footnote{Exceptions to this general feature are provided by the antenna complexes of photosynthetic bacteria that are adapted to survive under conditions of very weak illumination, e.g., green sulfur bacteria, which consist of very large numbers ($10^{3}-10^{4}$) of chromophores without a protein scaffold \cite{Prokhorenko:2000uc, Gan.Oos.etal-2009}.}.  This complexity indicates that a sophisticated understanding of how the nanoscale structure influences the light harvesting function is necessary in order to introduce precision and accuracy into the biomimetic approach.  In particular, recent studies of the behavior of the natural LHCs consisting of pigment-protein complexes have revealed evidence of quantum dynamical effects in the electronic energy transfer through the complex, indicating that quantum effects hitherto neglected in analysis of energy transport may play a role in the high quantum efficiency and need to be accounted for \cite{Che.Fle-2009}.  Theoretical modeling of the optical and electronic properties of large-scale pigment protein structures is a key component of obtaining this mapping and building a deeper understanding of the structure-function relationships in light harvesting.  

Much recent theoretical analysis has focused on understanding the quantum efficiency of energy transfer in biological LHCs -- e.g. Refs. \cite{Ola.Lee.etal-2008, Ish.Fle-2009b, Reb.Moh.etal-2009a, Car.Chi.etal-2009, Wu.Liu.etal-2010, Sar.Che.etal-2009, Mohseni:2011wg}. Detailed modeling, including incorporation of quantum coherent effects such as exciton delocalization and chromophore-protein interactions, has allowed rationalization of the energy transfer times seen in experiments and the high quantum efficiencies typical of LHCs.  Theoretical simulations have identified the dynamical parameters necessary for optimal energy transfer in model LHC systems and shown that a delicate balance between quantum coherent and incoherent dynamics appears to characterize this optimal energy transport regime \cite{Reb.Moh.etal-2009, Ple.Hue-2008}. Calculations exploring the landscape of these parameters for a small, well characterized LHC, the Fenna-Matthews-Olson (FMO) complex, have further indicated that this component of the green sulfur bacteria photosynthetic system operates in just such an optimal regime \cite{Wu.Liu.etal-2010, Mohseni:2011wg}. This would suggest that the design of optimized biomimetic LHCs should engineer the nanoscale structure so as to achieve strong inter-pigment couplings that compete with decoherent processes such as vibrational relaxation.   Since strong interpigment coupling is an important requirement for quantum coherence in electronic energy transfer, a related underlying question for the biomimetic LHC program is whether such quantum coherence is coincidental or whether it is essential to the light harvesting function. 

In this paper we take a different approach to the study of optimality in light harvesting by incorporating an important perspective from the field of multi-objective optimization.  Our starting point is the recognition that light harvesting, whether by natural or artificial systems, is not uniquely focused on achieving a high quantum efficiency for conversion of light to charge carriers, but is inherently a multi-objective optimization with several key objectives that must be simultaneously taken into account. In biological systems there are a wide array of objectives, not all of which necessarily have the same weight or status.  One could say that the most important objective for a biological system is survival, which will ultimately dictate the changes in design features in response to environmental changes.  In constructing biomimetic LHCs, we have a simpler task in that while the number of objectives are still greater than one, they are probably fewer and more equivalent in rank than in the biological case. For example, for photovoltaic applications one will likely want not only to maximize energy transfer and free-carrier generation efficiencies, but also to maximize the spectral width of absorption.  In contrast, for a sensor the second objective may be to maximize the sensitivity to a particular wavelength rather than the spectral width of absorption.   It is well known that in the presence of such multiple objectives one can have competition between them, which results in the emergence of families of optimal solutions that negotiate the trade-offs between the competing objectives in different ways \cite{Marler:2004ha}. 

The questions posed in this work are thus two-fold.  First, are there fundamental trade-offs involved in light harvesting? Second, if so, how can these be negotiated by engineering the structure of light harvesting complexes? We address these questions here by explicit calculation of the simultaneous optimization of the two desiderata mentioned above for a light harvesting system that might be used for energy conversion; we seek to simultaneously optimize the efficiency of excitonic transport and the spectral width of absorption. We do this by studying a prototypical biomimetic light harvesting antenna and requiring that it absorb photons in as wide a spectral window as possible while also efficiently transporting the resulting excitation energy to regions of charge separation.  The ultimate limits of spectral width are dictated by choice of pigments; that is, pigment transition energies largely determine the width of absorption profiles. Therefore, one method for increasing the absorption profile width is to include pigments with as many different transition energies as possible. However, this creates an energetically disordered aggregate with typically reduced transport efficiency.  Thus there is a \textit{trade-off}, or competition, between the two objectives.  One way to negotiate this conflict is to use molecular aggregation;  by employing strong Coulombic coupling, aggregates of pigments can broaden or sharpen absorption profiles as exemplified by the classic H- and J-aggregates absorption profiles \cite{Kob-1996}. In this work we shall explore the extent to which this provides an effective technique for negotiating the trade-off between efficiency and spectral-width.

We pause to mention some previous work examining the optimization and design of molecular structure in the context of light harvesting. Fetisova conducted some early and far-sighted studies into the relationship between structure and function in light harvesting complexes and strategies for optimization of structure, \textit{e.g.} \cite{Fetisova:1985cv, Fetisova:jb, Novikov:2006iw}. Much of this work was conducted well before experimental evidence for dynamical quantum coherence and therefore considered classical models of light harvesting dynamics.  More recently, Fingerhut \etal have employed structural optimization to design synthetic centers for ultra-fast charge separation for artificial photosynthesis \cite{Fingerhut:2009dm}. The charge separation dynamics was modeled classically using Marcus theory, and notably, they examined a multi-objective optimization landscape and considered trade-offs between the quantum efficiency of charge separation and other objectives. Knoester and co-workers have examined the excitonic and optical properties of general cylindrical aggregates of identical chromophores (with and without disorder) in several works, including Refs. \cite{Did.Kno-2004,Kno-2006}. Finally Noy \etal \cite{Noy:2006vd}, and more recently Scholes \etal \cite{Scholes:2011iq}, have compiled summaries of insights gained from studying natural LHCs and discussed how they aid the design of artificial light harvesting systems.

The remainder of this paper is structured as follows. Section \ref{sec:chrom_ass} introduces the physical systems forming the cylindrical assemblies of chromophores that constitute the focus of our study. Section \ref{sec:models} then outlines our theoretical models of the structure and electronic excitation dynamics in these systems. Section \ref{sec:results} presents the results of our multi-objective optimization studies in terms of achievable objectives and tradeoffs encountered. 
This is followed in section \ref{sec:str_params} by a detailed analysis of the structural and excitonic properties of the chromophore assemblies that optimize the objectives. Then in section \ref{sec:design} the preceding study and analysis is condensed into a set of design principles for light harvesting complexes, particularly cylindrical antennas. Finally, we conclude with a discussion and an outline of future work in section \ref{sec:conclusions}.
 
\section{Cylindrical chromophore assemblies}
\label{sec:chrom_ass}
We focus in this work on cylindrical molecular assemblies as a prototypical architecture for biomimetic LHCs. Many systems self-assemble or can be templated into cylindrical structures and several of these have been studied as candidates for artificial light harvesting systems. Examples include carbocyanine molecules with hydrophobic and hydrophilic side groups and porphyrin derivatives that aggregate into cylindrical structures  \cite{Eisele:2009cf,Vlaming:2011ib}. There are also examples of cylindrical molecular aggregates in natural LHCs, the most prominent one being the \textit{chlorosome} complexes of green sulfur bacteria \cite{Prokhorenko:2000uc, Gan.Oos.etal-2009}. 

One particularly promising realization of a cylindrical aggregate is not formed from the direct aggregation of chromophores, but rather, from \textit{virus-templated assembly}. Virus-templated chromophore assemblies are supramolecular complexes constructed by attaching chromophores to protein coats of viruses that then self-assemble into large, regular structures. The self-assembling protein coats are used as rigid scaffolds that guide the synthetic organization of chromophores. Structures templated using the tobacco-mosaic, M13 and MS2 viruses, amongst others, have been demonstrated for use in light harvesting \cite{Mil.Pre.etal-2007, End.Fuj.etal-2007, Nam:2010fw}, drug delivery \cite{Wu.Hsi.etal-2009}, and battery technologies \cite{Nam.Kim.etal-2006,Roy.Gho.etal-2008}. This method of templated self-assembly presents a particularly promising path towards producing engineered assemblies of chromophores with controlled nanoscale structure. The templated assembly allows for a greater degree of customization than direct aggregation of chromophores because the nanoscale structure can be controlled by the choice of chromophores, templating protein, and of chromophore-protein linker groups. Templated assembly has the further advantage that pigments which do not naturally self-aggregate can be used. Also, it should be noted that the virus-templated assembly process creates a protein-pigment structure and not a direct molecular aggregate; as noted above, the former are more common in natural LHCs.

\begin{figure}[ht]
\centering
\subfigure[~Templated assembly using TMV]{
\includegraphics[scale=0.32]{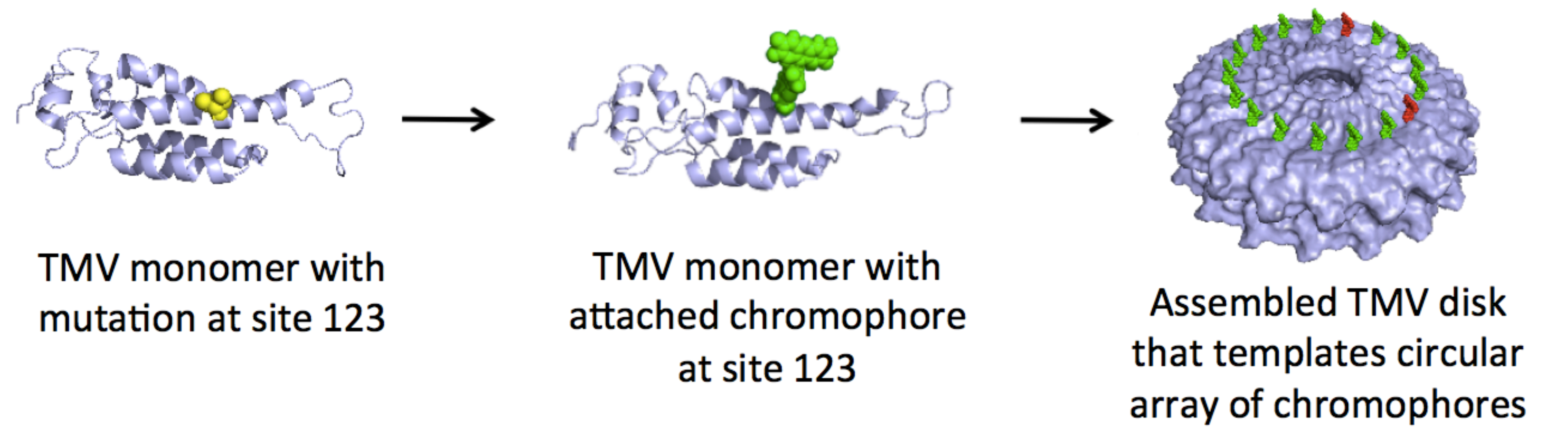}
\label{fig:TMVa}
}
\subfigure[~Variety of sites on a TMV protein monomer at which mutations can be introduced to facilitate covalent attachment of chromophores]{
\includegraphics[scale=0.4]{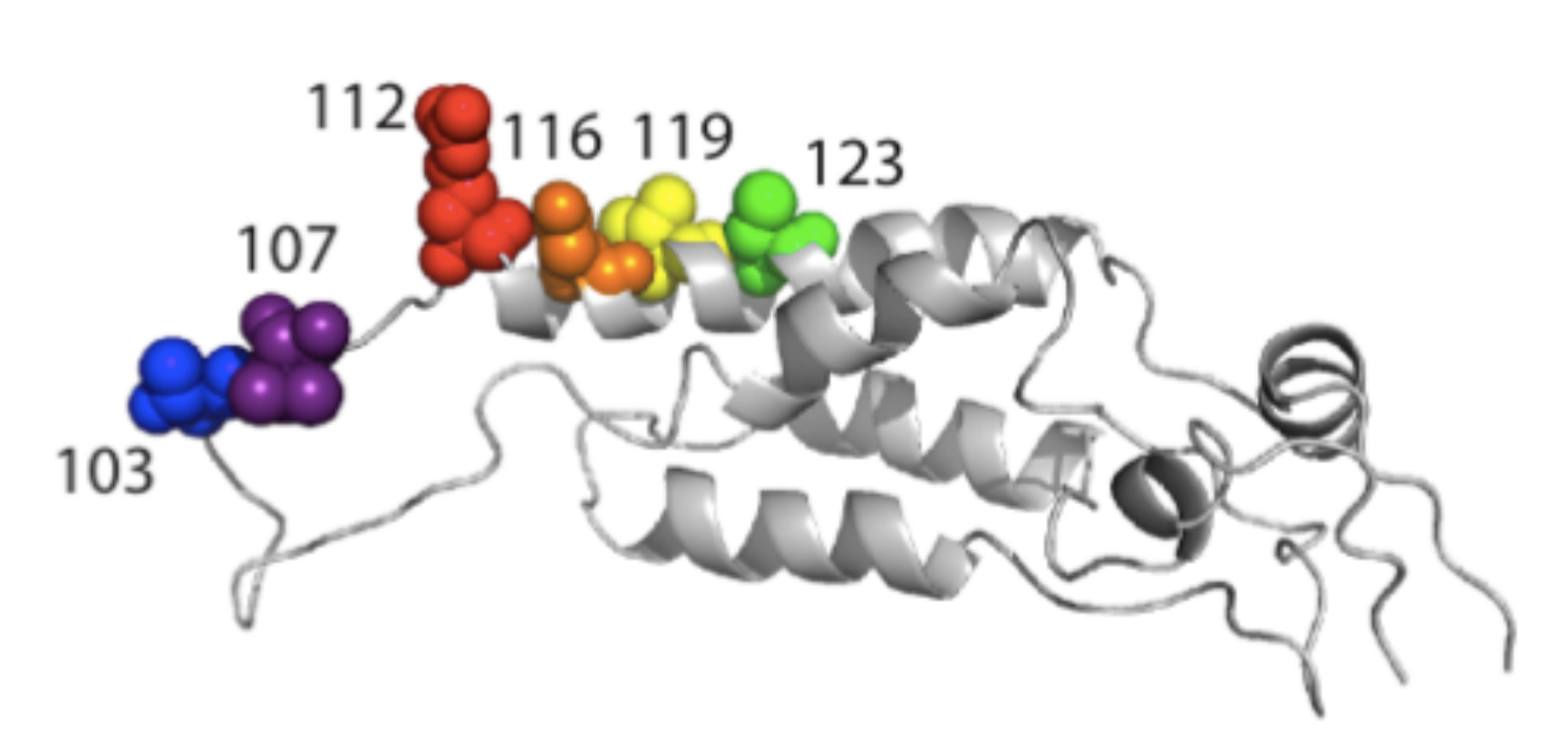}
\label{fig:TMVb}
}
\caption{The self-assembling TMV protein provides a scaffold for producing an array of chromophores with well defined inter-chromophore distances.
\label{fig:TMV}}
\end{figure}

We consider here templated assemblies based on use of the tobacco mosaic virus (TMV) as a scaffold.  The protein of this virus can self-assemble into stacked disks or cylinders, depending on assembly conditions such as solution pH \cite{Klu-1999}. From a practical stand-point TMV assemblies are advantages because the assembly process is very well known and large quantities can be reliably produced. Various laboratories have demonstrated covalent attachment of pigments to various sites on mutated TMV protein monomers \cite{Mil.Pre.etal-2007, End.Fuj.etal-2007} (see Fig. \ref{fig:TMV}), and hence shown that the inter-pigment distances can be tuned by choice of attachment site. This is advantageous for studying the impact of quantum coherent effects on light harvesting, because coherent dynamics will be more prevalent in densely packed structures where inter-pigment couplings are strong.  Thus the ability to vary the inter-pigment distance controllably over a finely tuned and large range of values, with other features of the protein scaffold held constant, allows the synthesis of chromophore arrays with and without the potential for such coherent dynamics.

\section{Structural and dynamical model}
\label{sec:models}
\subsection{Structural model}
\begin{figure}[htbp]
\begin{center}
\includegraphics[width=12cm]{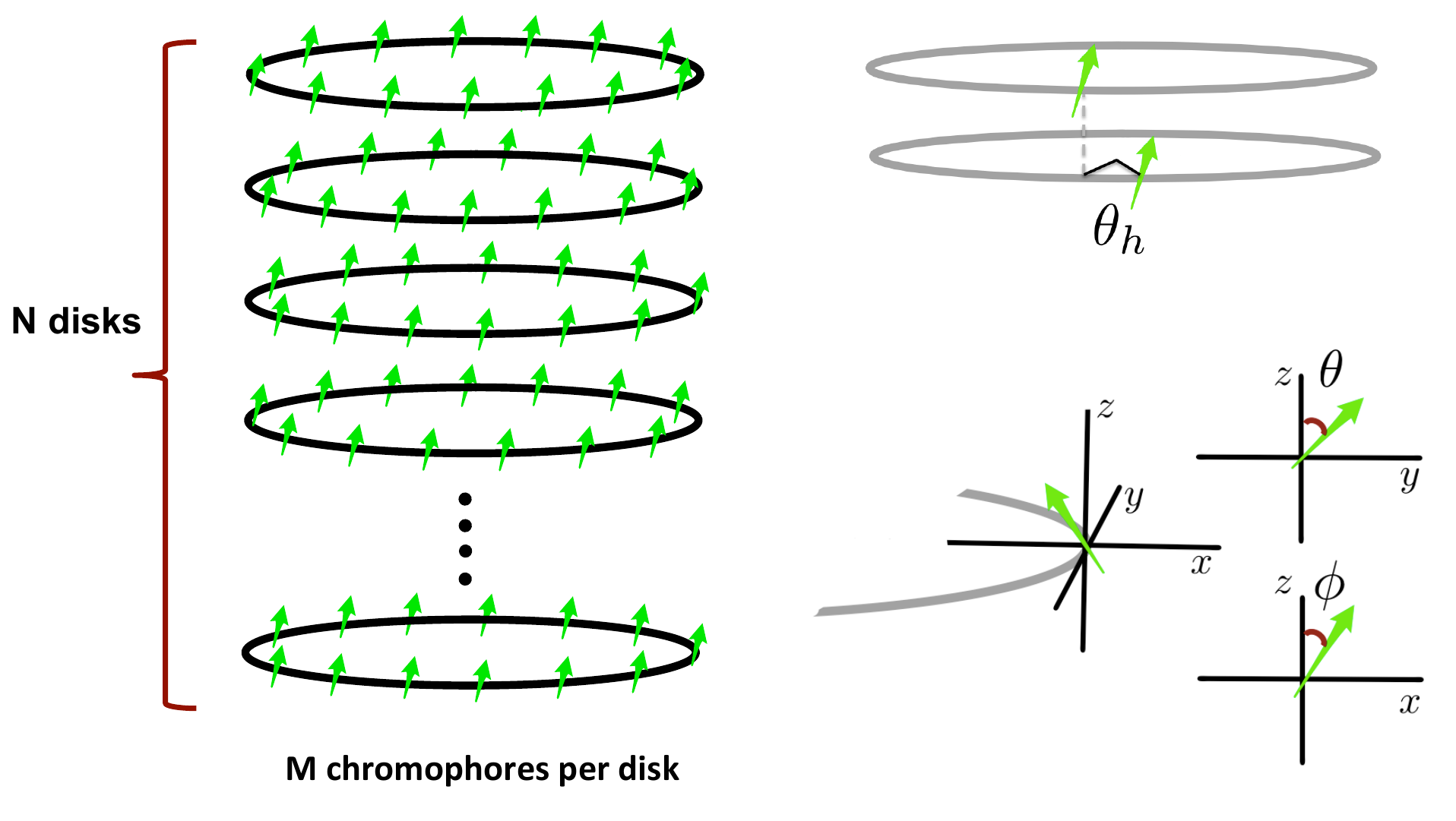}
\caption{Reduced description of chromophore assembly resulting from stacked disk TMV structure, in which only the chromophore properties are variable and the TMV protein scaffold is held constant. There are $M$ chromophores attached to each disk and $N$ disks. All chromophores are assumed to be attached with the same orientation and this orientation defines a transition dipole for each chromophore which is specified by two angles, $\theta$ and $\phi$. Finally, an angle $\theta_h$ specifies the degree of misalignment between neighboring disks, and allows for a helical twist.  \label{fig:chrom_array}}
\end{center}
\end{figure}

Our multi-objective optimization studies are made for an idealized but versatile model for cylindrical arrays of chromophores that is shown in Fig. \ref{fig:chrom_array}. The model consists of $N$ disks, each with $M$ chromophores attached at specific sites, as determined by an implicit protein scaffold or template.  Chromophore location on sequential disks may be off-set by a variable amount and the chromophores have variable orientation but they are restricted to all have the same orientation relative to their disk.  This model allows us to simulate TMV-templated chromophore assemblies in cylindrical and disk-like structures, which may have vertical or helical stacking of chromophores along the cylinder. The cylindrical array of chromophores will function as an antenna absorbing photons and transporting the resulting photo-excitation to one of its ends. This end of the cylinder (referred to as the ``bottom" of the cylindrical aggregate) interfaces with a surface or electron acceptor and charge separation occurs at this interface. For the TMV-templated structures a distinct chromophore whose energy levels match well with the remainder of the chromophores and the surface work function can be attached at the end to facilitate the electron transfer event.  In this first work, we will not be concerned with optimizing these subsequent charge separation and transfer events but rather with the optimization of the preceding energy transfer to the separating interface. 

In the present calculations we assume that the following  degrees of freedom (design parameters) can be tuned: (i) the transition energy of the chromophores on each disk (i.e., each of the $N$ disks has a distinct energy, but the $M$ chromophores on any single disk have the same transition energy), (ii) all chromophores attach to the TMV protein scaffold with the same orientation and this orientation defines a transition dipole that is specified by two angles, $\theta$ a tangential angle and $\phi$ a radial angle, (iii) there can be a degree of misalignment between neighboring disks and this defines a helical angle, $\theta_h$, which allows introduction of a helical twist along the cylinder coordinate.  See Fig. \ref{fig:chrom_array} for representations of these angles. This results in a total of $N+3$ controllable degrees of freedom, which may be used to produce a wide range of chromophore array structures that range from complete alignment along any axis, to helical arrays with variable numbers of helical strands, e.g., $M=1$ with $\theta_h > 0$ corresponds to a single helical strand, $M=2$ with $\theta_h > 0$ to a double helix, etc. TMV can template a 1-helix at the appropriate conditions, however we will not study this structure in this work, and will instead focus on $M=17$ as specified below. In addition, we note that the chlorosome of green sulfur bacteria is assumed to have a structure that is an $M$-helix \cite{Gan.Oos.etal-2009} and therefore this model is  flexible enough to capture artificial and natural light harvesting structures.

The calculations presented here make explicit comparison of two TMV-templated aggregates based on the stacked disk morphology, which are designed to have very different inter-chromophore distances.   Upon aggregation each TMV ``disk" consists of 17 protein monomers. Assuming complete functionalization (attachment of chromophores) this results in $M=17$ chromophores per disk. Each chromophore is attached to the disk at a specific location, which is determined by its binding site on a protein monomer: this can be varied by prior treatment of the protein, allowing variable attachment location and hence variable inter-chromophore distance, see Fig. \ref{fig:TMV}.  We compare here TMV103 and TMV123, where the numerical labels refer to the protein monomer site where pigments are attached. The 103 site is within the pore of the assembled TMV protein and attachment here results in a disk of chromophores of radius $25$~\AA, implying an average neighboring pigment separation distance (within a disk) of about $10$~\AA, and thus strong coupling of chromophores. In contrast, attachment at site 123 results in a disk of chromophores of radius $40$~\AA~and an average neighboring pigment separation distance of $14-15$~\AA, which leads to weaker electronic couplings. The vertical distance between disks remains the same for both structures: $\sim 20$~\AA. While TMV103 can achieve a greater density of pigments, we note, however, that the coupling between pigments on neighboring disks can nevertheless still be the same for both structures because the distance between disks is the same. A question we seek to answer in the following analysis is whether the dense packing within disks and any resultant increase in coherent dynamics of TMV103 is advantageous for any aspect of light harvesting.

We note that current experimental techniques do not allow independent tuning of all the above design parameters for TMV-templated cylindrical assemblies. For instance, it is difficult to restrict a disk to only containing chromophores of a given species (transition energy) when there are several species on the whole cylinder. However, it is important to choose a large number of independent degrees of freedom in order to explore the landscape of optimal light harvesting given near-ideal resources. The figures of merit obtained can therefore be considered upper bounds on what is currently experimentally achievable. In a forthcoming publication we will examine the optimization landscape for a more modest set of design parameters.

\subsection{Dynamical model}
The TMV-templated cylindrical chromophore assemblies are complex biomolecular systems. In order to efficiently compute and compare their light harvesting properties we must resort to an effective description of the electronic degrees of freedom. In analogy with the conventional modeling of natural light harvesting complexes \cite{Che.Fle-2009,Ish.Cal.etal-2010} we employ the Frenkel Hamiltonian to describe coherent dynamics:
\beq
H_e = \sum_i E_i \ket{i}\bra{i} + \sum_{i,j}J_{ij} (\ket{i}\bra{j} + \ket{j}\bra{i})
\eeq
where $\ket{i}$ denotes an electronic excitation localized on pigment $i$ in the complex, $E_i$ is the transition energy of pigment $i$ and $J_{ij}$ denotes the Coulombic coupling of pigments $i$ and $j$. For simplicity, we assume that each pigment has only one dominant transition in the wavelength region of interest. This model can be easily generalized to pigments with more than one transition. In this work we treat the electronic coupling in the dipole-dipole coupling approximation. This approximation is generally valid for inter-pigment separations greater than $\sim 12 \AA$ \cite{MunozLosa:2009dl, Yang:2010kg}, and while we will model some cases where the inter-pigment distance is slightly smaller than this, the error due to the dipole-dipole approximation in those cases will not affect our conclusions significantly. The eigenstates of $H_e$ are the excitonic states in the complex, whose decomposition in terms of the site basis $\{ \ket{i} \}$ we denote:
\beq
\ket{e_k} = \sum_i U_{i,k} \ket{i}
\label{eq:basis_tr}
\eeq
where $U$ is the matrix that diagonalizes $H_e$. In addition to this coherent dynamics, the coupling of pigment electronic degrees of freedom to protein degrees of freedom, and the resultant decoherent and dissipative dynamical effects must be incorporated into the simulation. There is little experimental information on the vibrational dynamics of the protein degrees of freedom in virus-templated assemblies and hence we employ a model that makes minimal assumptions. The electronic-vibrational coupling is chosen to be linear and the vibrational degrees of freedom are modeled harmonically, resulting in the interaction Hamiltonian:
\beq
H_I = \sum_i \ket{i}\bra{i} \sum_\xi c_{\xi,i}(a_{\xi,i} + a\dg_{\xi,i})
\eeq
where $a_{\xi,i}$ denotes the annihilation operator for mode $\xi$ coupled to the excited state of pigment $i$, and the vibrational ``bath" is described by a free harmonic Hamiltonian: $H_v = \sum_{i,\xi} \hbar \omega_{\xi,i}a\dg_{\xi,i}a_{\xi,i}$. We model the harmonic bath using an over-damped Brownian oscillator model \cite{Muk-1999} in the high temperature limit. The spectral density is taken to be Ohmic with Lorenz-Drude regularization: $J(\omega) = \frac{2\lambda \gamma \omega}{\omega^2 + \gamma^2}$, with reorganization energy $\lambda = 100 \textrm{cm}^{-1}$, and relaxation time $\gamma^{-1} = 100$fs. 
 The true vibrational and solvent dynamics of TMV-templated assemblies likely has a more complex spectral density than this structureless model. However, in the absence of experimental data we use this simple spectral density with minimal assumptions, noting that its generic features and parameter values are similar to the spectral densities used for simulation of electronic energy transfer in natural LHCs \cite{Ish.Cal.etal-2010}. 

In order to describe the effective dynamics of the electronic degrees of freedom that are of interest, we average over the harmonic bath degrees of freedom. There are numerous methods for doing this averaging, at various levels of approximation. In this work we use the modified Redfield formalism \cite{Yan.Fle-2002,Zha.Mei.etal-1998} to derive an effective equation of motion for the excitonic populations of the complex once the vibrational degrees of freedom have been averaged.  In effect the modified Redfield approach derives a dynamical equation for the exciton populations:
\beq
\frac{\textrm{d}P(t)}{\textrm{d}t} = R P(t)
\eeq
where $R$ is the modified Redfield rate matrix and $P(t)$ is a vector of excitonic populations of length $n + 2$, where $n=N\times M$ is the total number of chromophores in the complex, and we have included two additional entries to track population lost via exciton loss (recombination) and excitation trapping (due to localization at the interface and subsequent charge separation). The exciton loss and trapping rates are chosen phenomenologically since there are no experimentally measured values for TMV-templated chromophore aggregates).  See \ref{sec:dyn_model} for details. An important feature of this approach is that while the dynamics of off-diagonal elements of the exciton density matrix, which correspond to the excitonic coherences, are not explicitly followed in time, the components of the interaction Hamiltonian that generate them are included perturbatively in the calculation of the population transfer rate matrix elements, see Eq. (42) of \cite{Yan.Fle-2002}. 

We employ the modified Redfield formalism here for several reasons. First, it has been shown to be reasonably accurate across a wide range of parameter regimes, varying from strong to weak system-bath coupling, and is thus preferred to the standard Redfield approach outside of the regime of weak system-bath coupling \cite{Yan.Fle-2002}.  The approach is particularly effective at modeling energy transfer in systems with structural or energetic disorder that is larger than, or comparable to, the magnitude of bath induced reorganization effects \cite{Zha.Mei.etal-1998}. Second, it is an extremely efficient method of dynamical propagation and this efficiency will be critical for the numerically intensive optimizations we undertake. The primary drawback of the modified Redfield model is that it only allows the explicit dynamical simulation of excitonic populations and not of coherences. Given that the excitonic populations will nevertheless be perturbatively influenced by the coherences (as described above), this is not a significant drawback for our purposes since we are primarily interested in asymptotic efficiency of transport and spectral width of absorption, both of which can be extracted from excitonic populations (see \ref{sec:dyn_model}). 

Finally, we note that using the symmetries of the cylindrical stacked disk structure together with the dipole-dipole approximation implies that we only have to consider a limited domain for the two angles that define the transition dipole of a chromophore: $\theta \in [0, \pi/2]$, $\phi \in [-\pi/2,\pi/2]$, as well as for the helical angle $\theta_h \in [-\pi/M, \pi/M]$. The Hamiltonians resulting from all other choices of these angles can be replicated by a choice within these domains -- e.g. the coupling between disks with all dipoles pointed up and all dipoles pointed down yields the same excitonic structure. Therefore, in order to improve the efficiency, we allow the angular design variables to vary within these restricted domains. 

\section{Optimized light harvesting in cylindrical chromophore assemblies}
\label{sec:results}

\subsection{Multi-objective optimization and Pareto fronts}
The multi-objective optimizations described here employ $N=10$ disks in the cylindrical aggregate, with the energy of the chromophores on any disk allowed to vary in the range $400 - 450$nm. We assume full functionalization of the TMV protein monomers and hence there are $M=17$ chromophores per disk. That is, TMV naturally assembles into disks with $17$ protein monomers \cite{Mil.Pre.etal-2007}, and assuming each protein monomer is functionalized with a chromophore, this results in $M=17$ chromophores per disk.  All chromophores are assumed to have the same transition dipole strength, $|\mu|=3$ Debye. When the two angles defining the orientation of the transition dipole of the chromophore, $\theta, \phi$ and the helical angle $\theta_h$, are also included, this yields $13$ design parameters that define the Hamiltonian of the electronic degrees of freedom of the chromophore assembly. We seek to optimize over this design parameter space in order to find the structures that are optimal for two objectives: transport efficiency and spectral width of absorption.  The maximization of these objectives is performed here using a genetic optimization algorithm that evaluates the energy transfer efficiency and absorption spectral width for each member of its evolving population. \ref{sec:mo_opt} provides more detail on the specifics of the genetic optimization algorithm. Assemblies with and without disorder are considered, where the disorder can be both structural and energetic. For the simulation of systems with disorder, we average the objectives over $504$ instances of disorder at each design parameter configuration. The disorder is introduced as independent, Gaussian distributed perturbations of the relevant parameters, i.e., of each chromophore's transition energy (with variance $2$nm), transition dipole orientation angles $\theta, \phi$ (with variance $0.2$ radians), and helical angle $\theta_h$ (with variance $0.2\times 2\pi/M$ radians). Note that the introduction of disorder will break the equivalence of the $M$ chromophores on a single disk.

In the field of multi-objective optimization,  a critical concept is the \textit{Pareto front}. This is the locus or curve in objective space formed by all the solutions to the optimization program that optimally negotiate the trade-offs between multiple objectives, where this means that increasing the value of a particular objective (for a maximization program), leads to the decrease in the value of one or more of the other objectives. The shape of the Pareto front reveals the amount of competition between the various objectives. In cases where there is significant competition between different objectives, one must choose a trade-off solution that suitably compromises between the competing objectives. See \ref{sec:mo_opt} for more details on Pareto fronts and multi-objective optimization. Our genetic optimization is designed to converge on the Pareto front for the problem of light harvesting using a cylindrical antenna, with the dual objective functions of efficiency of excitonic transport and spectral width of absorption.

\subsection{Pareto fronts for TMV103 and TMV123}
\begin{figure}[htbp]
\begin{center}
\includegraphics[width=11cm]{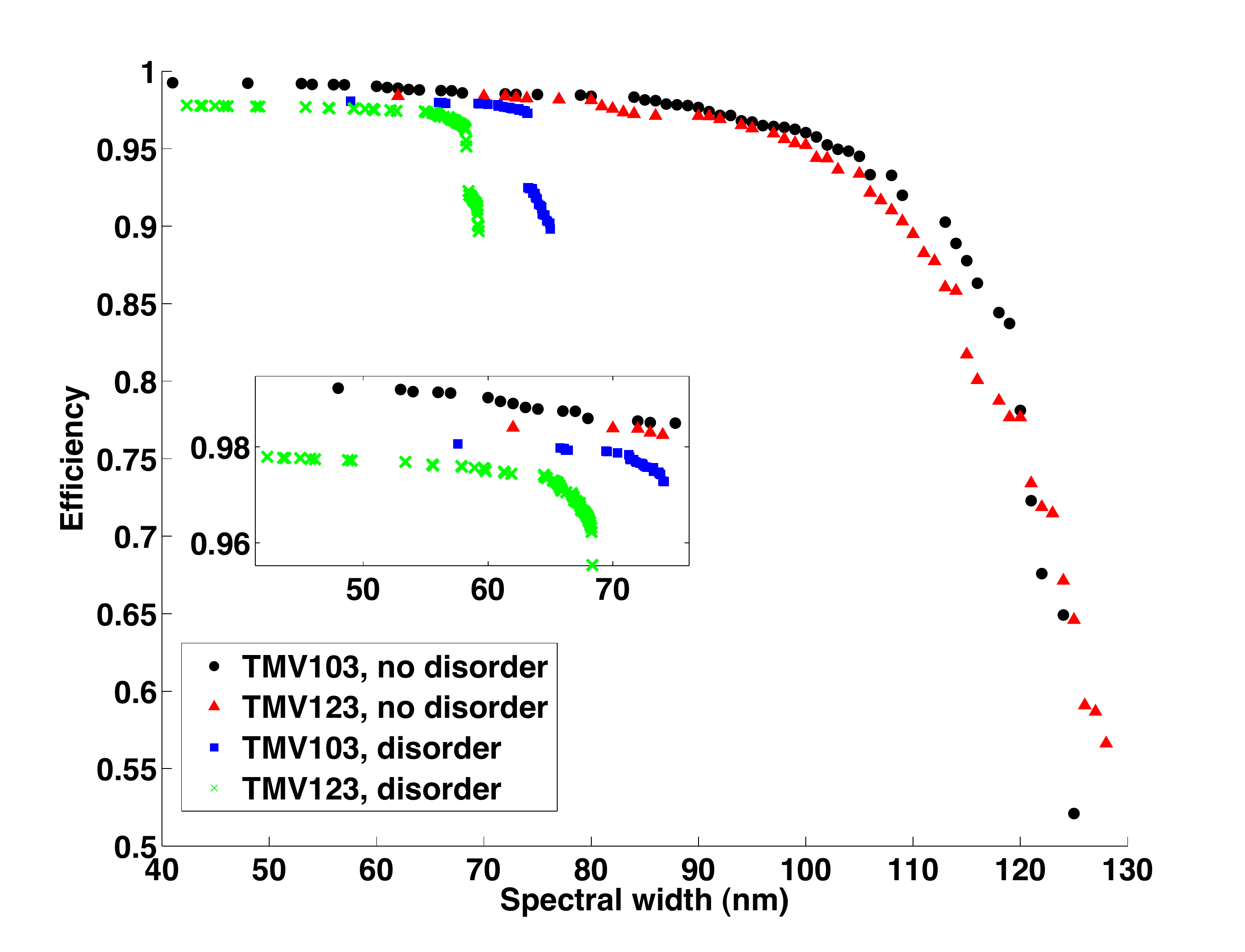}
\caption{Pareto fronts in the efficiency-spectral width objective space for TMV103 and TMV123 structures with and without structural and energetic disorder. The statistical variation in the points with disorder is negligible (we confirmed this) at the level of sampling performed and therefore error bars have not been included for clarity. The inset is a zoom into the congested high efficiency region. \label{fig:pareto_width}}
\end{center}
\end{figure}

The Pareto fronts for simultaneous optimization of spectral-width and efficiency for TMV103 and TMV123 are shown in Fig. \ref{fig:pareto_width}, for both calculations with and without disorder,  First consider the fronts for the ideal structure in the case of no disorder (black dots and red triangles for TMV103 and TMV123, respectively). The efficiencies and spectral widths achievable are very similar for both the densely packed (TMV103) and sparsely packed (TMV123) structures, and the shape of the trade-off curve is very similar too. The presence of curvature in both of these Pareto fronts indicates that there is a trade-off between achieving high spectral width and high efficiency. In particular, we see that for low spectral width absorption, both of these idealized cylindrical structures can achieve near unit efficiency of electronic energy transport.  The similarity of these two fronts at all except the largest spectral widths suggests that in the absence of disorder there is generally little advantage achieved by dense packing and strong coupling of pigments; the same efficiencies and spectral widths can be achieved with dense and sparse packing.  The exception is the regime of high spectral width, where there appears to be a crossover to and from a regime in which the densely packed TMV103 shows a better simultaneous optimization of the two objectives, consistently achieving a higher efficiency for a given spectral bandwidth (in the range 100-120 nm),  In general, however, the results without disorder show that the negotiation of the tradeoff between the two objectives is comparable for densely and sparsely packed structures.

With the introduction of structural and energetic disorder, several marked changes occur to the Pareto fronts. Firstly, both fronts collapse in the horizontal direction, reflecting the result that much smaller spectral widths are now achievable. There is also a small but noticeable amount of collapse in the vertical direction, indicating that somewhat smaller efficiencies are achievable with disorder.  In addition, the Pareto fronts for TMV103 and TMV123 separate and become distinguishable. These generic changes can be rationalized in term of energetic and structural features, which we will do below. For low spectral width absorption, the densely packed TMV103 (blue squares) shows a slight enhancement in efficiency over the sparsely packed TMV123 (green crosses), although both are still very close to the optimal values obtained with the ordered structures.  It is noteworthy that the best efficiencies achievable are comparable for TMV123 and TMV103, and somewhat surprisingly, not very different from the best efficiencies achieved in the absence of disorder. This indicates that efficiency of energy transfer in cylindrical structures can be robust to disorder. In contrast to this robust behavior of the efficiency, the achievable spectral widths are however now much smaller in the presence of disorder.  Here the dense packing of TMV103 is clearly beneficial and allows significantly greater spectral width to be attained. In general terms, the overall shapes of the Pareto fronts in the presence of disorder are similar for TMV103 and TMV123, and show that a more drastic trade-off between efficiency and spectral bandwidth is required than in the absence of  disorder.  Thus, to increase spectral width beyond $\sim 68$nm ($\sim 75$nm) for TMV123 (TMV103) requires a significant sacrifice in efficiency of energy transport. 

The generic nature of the Pareto fronts obtained for the two TMV-templated chromophore assemblies with and without disorder indicate that the trade-off between efficiency and spectral width is fundamental for such cylindrical light harvesting antennas and independent of the presence of disorder.  Beyond this result however, several subtle differences are apparent between the detailed behavior of the densely and sparsely packed systems.  In particular, the densely packed TMV103 appears to be better able to preserve both a high efficiency and high absorption bandwidth in the presence of disorder.  This suggests that the dense packing and ensuing quantum coherence may serve to provide robustness of function against sources of disorder.  In the following section, we explore these factors in more detail by examining the parameters defining the structures on the Pareto fronts. 

\section{Design variables emerging from optimal structures}
\label{sec:str_params}
\begin{figure}[]
\centering
\subfigure[~TMV103 without disorder]{
\includegraphics[scale=0.44]{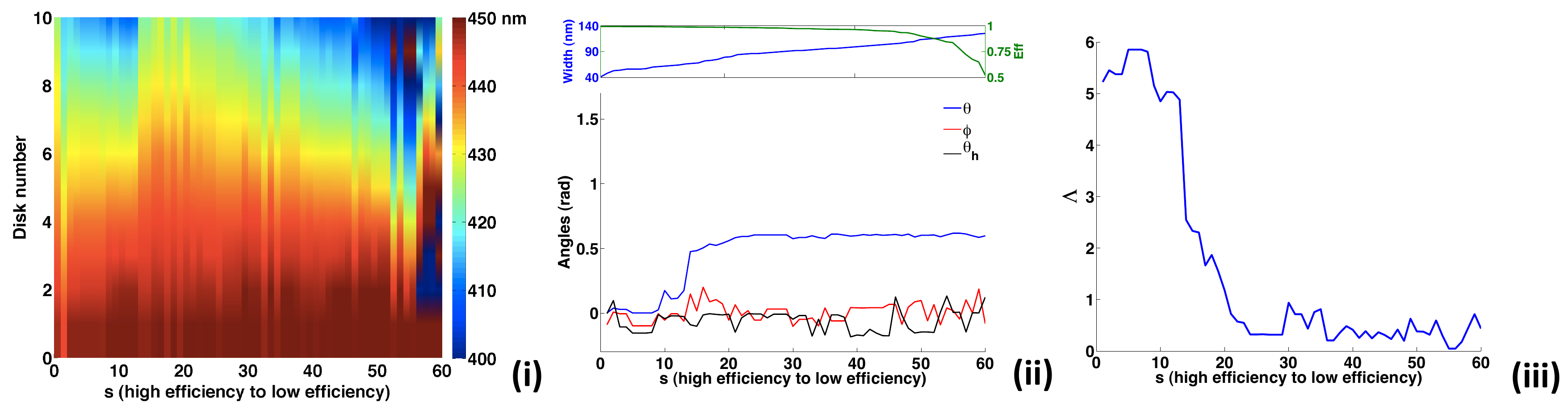}
\label{fig:wp_1}
}
\subfigure[~TMV123 without disorder]{
\includegraphics[scale=0.44]{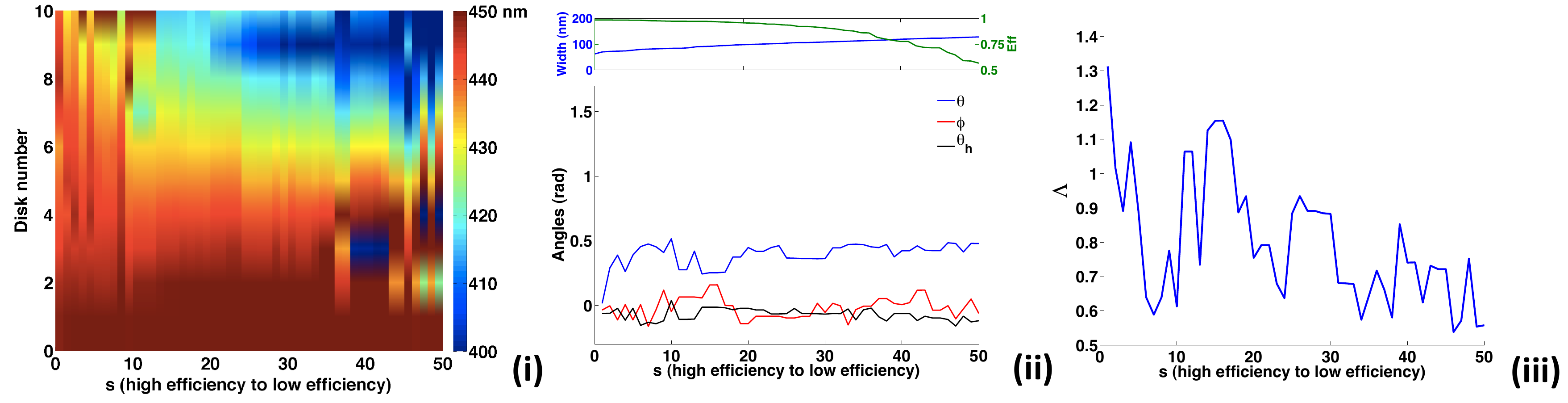}
\label{fig:wp_2}
}
\caption{Design variables for structures on the Pareto fronts \textit{without disorder}, for (a) TMV103 and (b) TMV123. The $x$-axis on all plots shows the structure index $s$ on the Pareto front. Left panels (i): color coded disk energies for the $N=10$ disks. Central panels (ii): transition dipole orientation angles $\theta, \phi$ and helical angle $\theta_h$, with the upper subplot showing the variation of efficiency (green line) and spectral width (blue line) as a function of $s$. Right panels (iii): ratio of intra- to inter-disk dipole-dipole coupling, $\Lambda$. 
\label{fig:parameters_nodisorder}}
\end{figure}

\begin{figure}[ht]
\centering
\subfigure[~TMV103 without disorder]{
\includegraphics[scale=0.58]{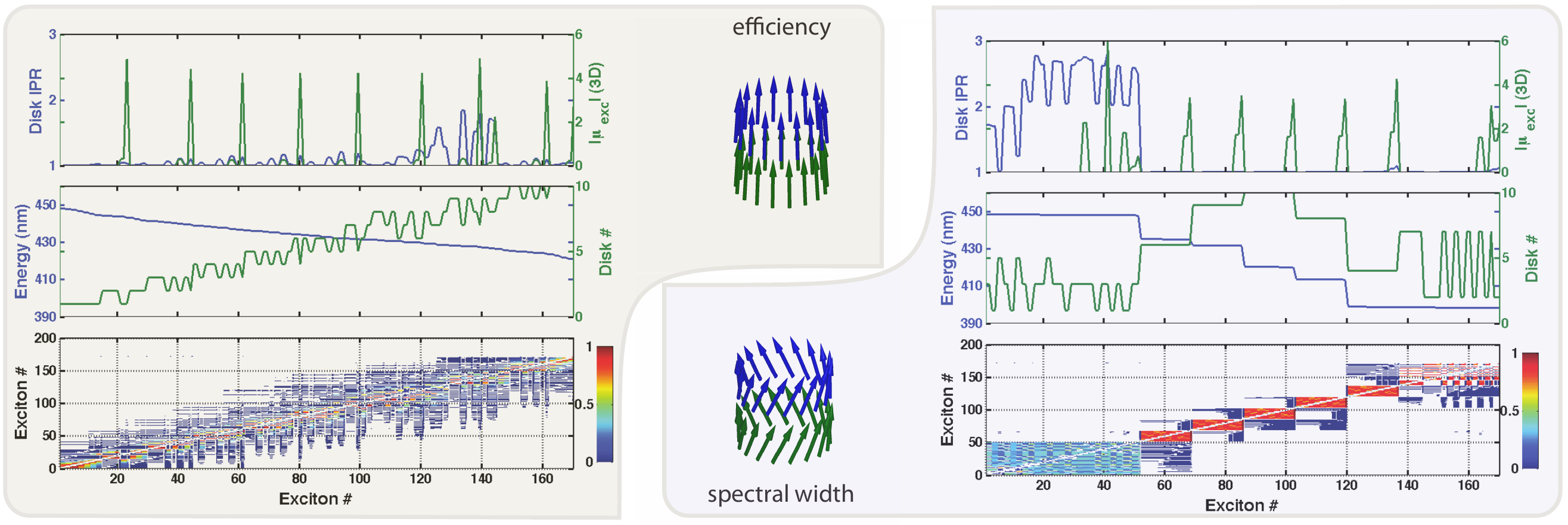}
\label{fig:ex_1}
}
\subfigure[~TMV123 without disorder]{
\includegraphics[scale=0.58]{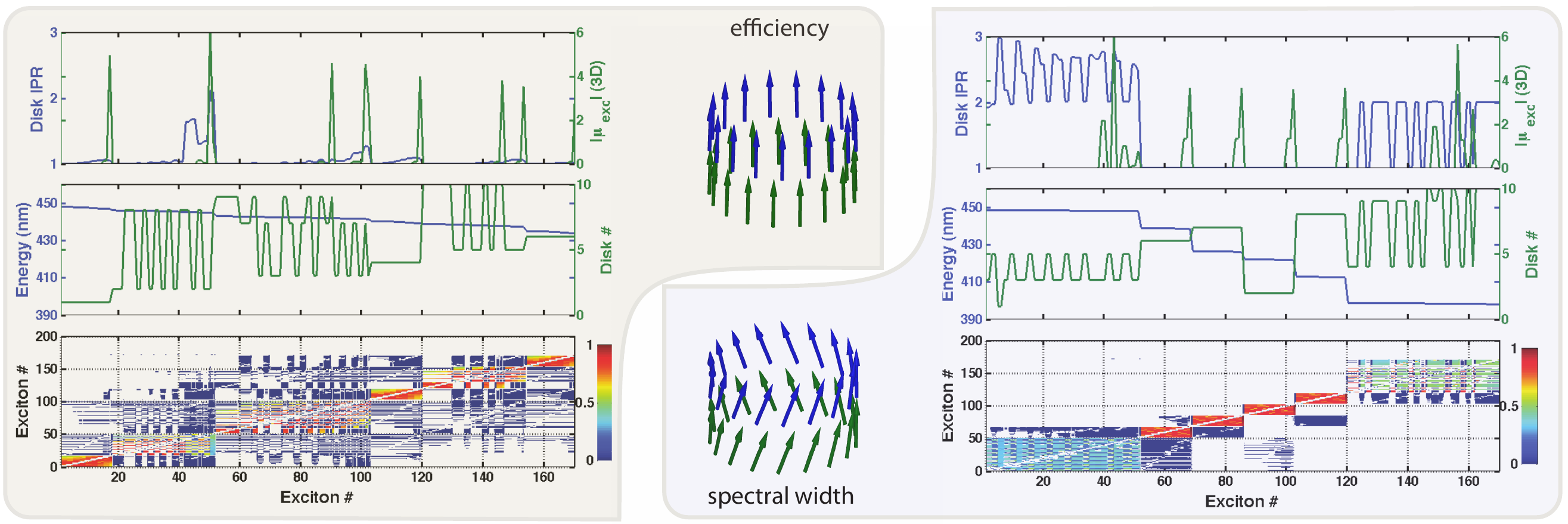}
\label{fig:ex_2}
}
\caption{
Analysis of excitonic structure and dynamics for the extremal structures on the TMV103 and TMV123 Pareto fronts (\textit{without disorder}). See main text (beginning of section \ref{sec:str_params}) for explanation of the plots. 
\label{fig:extremes_nodisorder}}
\end{figure}

To understand the Pareto fronts in Fig. \ref{fig:pareto_width} and to appreciate how the objectives are maximized and the conflicts negotiated, we now examine the performance metrics in parameter space. We use two complementary summaries of the structural information and present them in Figs. \ref{fig:parameters_nodisorder} and \ref{fig:extremes_nodisorder} for the case of no disorder, and Figs. \ref{fig:parameters_disorder} and \ref{fig:extremes_disorder} for the case with disorder. The first summary in both cases (Figs. \ref{fig:parameters_nodisorder} and \ref{fig:parameters_disorder}) presents the parameter variations for structures on the Pareto front, while the second (Figs. \ref{fig:extremes_nodisorder} and \ref{fig:extremes_disorder}) presents a detailed analysis of the excitonic and dynamical features of extremal structures on the Pareto fronts in Fig. \ref{fig:pareto_width}. In the next two subsections we will interpret the data in these summaries, but first we describe in detail the content of the figures.

Figs. \ref{fig:parameters_nodisorder} and \ref{fig:parameters_disorder} summarize the values of the 13 design parameters for structures on the Pareto front, for TMV103 and for TMV123 without (Fig. \ref{fig:parameters_nodisorder}) and with (Fig. \ref{fig:parameters_disorder}) disorder. The $x$-axis on all plots in these figures varies over the integer parameter $s$, which indexes structures on the Pareto fronts in Fig. \ref{fig:pareto_width} from left to right, i.e., starting with the most efficient (smallest spectral width) and ending with the least efficient (largest spectral width).  There are three panels of plots displayed in each subfigure of Figs. \ref{fig:parameters_nodisorder} and \ref{fig:parameters_disorder}.  
The left panels, (i) show a color-coded representation of the energies of the $N=10$ disks for each structure on the corresponding Pareto front.  The middle panels, (ii) show the two orientational angles $\theta, \phi$ and the helical angle $\theta_h$ for each structure on the corresponding Pareto front. The third panels, (iii) show the ratio of coupling between neighboring chromophores on a single disk (intra-disk) to the coupling between closest neighboring chromophores on adjacent disks (inter-disk), which we define as $\Lambda \equiv J_\textrm{\scriptsize intra-disk}/J_\textrm{\scriptsize inter-disk}$. This ratio is important because it indicates the direction of strong electronic coupling -- either within disks (large values of the ratio) or between disks and along the cylinder (small values of the ratio). It is important to appreciate that the electronic coupling ($J$) by itself does not dictate the rate of energy transfer between two chromophores (or a group of chromophores). The energy difference between them also plays a role. However, the value of $\Lambda$ informs us about the direction of dominant coupling and thus will be used in the interpretation of results below. 

Figs. \ref{fig:extremes_nodisorder} and \ref{fig:extremes_disorder} present the excitonic structure and dynamics of the extremal structures on each of the four Pareto fronts shown in Fig. \ref{fig:pareto_width}. The extremal structures are defined as the solutions on each of the Pareto fronts that achieve the maximum of one of the objectives at the expense of the other: these are precisely the structures indexed by minimal and maximal values of $s$. Examining these extremal structures elucidates the design principles involved in maximizing the objectives. Fig. \ref{fig:extremes_nodisorder} examines extreme structures in TMV103 and TMV123 in the absence of disorder, and Fig. \ref{fig:extremes_disorder} examines extremes structures for TMV103 and TMV123 in the presence of disorder.
Each subfigure in these plots is divided into two panels; the left panel shows details of the most efficient structure and the right panel shows details of the structure that achieves the greatest spectral width. Between these two panels, in the center of the figure we show the corresponding optimal orientations of the chromophore transition dipoles in the two extremal structures.  Note that these orientations are shown for only two of the $N=10$ disks, the orientations on the remaining disks are identical to these. The structure giving the most efficient transport is on top and is to be viewed together with the left panel.  The structure giving the greatest spectral bandwidth is below and is to be viewed together with the right panel.  

Considering now the left and right panels of Figs. \ref{fig:extremes_nodisorder} and \ref{fig:extremes_disorder}, each of these shows three plots. In all of these plots the $x$-axis indexes the excitons (sorted from lowest to highest energy) for the corresponding extremal structure.  Various properties that quantify the delocalization and energetics of the excitons and the resulting modified Redfield rates for energy transfer are shown as a function of the exciton index, as follows.
\begin{enumerate}
\item The top plot shows two measures that quantify the optical accessibility and delocalization of the excitons, respectively. The green line shows the magnitude of the excitonic dipole in units of the individual chromophore transition dipole magnitude ($3$ Debye), which constitutes a measure of the optical accessibility of the excitons. The blue line shows the inverse participation ratio (IPR) of the exciton between disks , defined as: $\textrm{dIPR}_k = \frac{1}{\sum_{n=1}^{10} {p^k_n}^2}$ where $p_n^k$ is the probability of exciton $k$ being localized on disk $n$ which is a sum of the probabilities of exciton $k$ being localized on any of the chromophores composing disk $n$: $p^k_n = \sum_{i} |U_{i,k}|^2$ where the sum is over chromophores $i$ that compose disk $n$ and $U_{i,k}$ are elements of the transformation matrix that defines the site-basis to exciton-basis transformation, \erf{eq:basis_tr}. The disk inverse participation ratio dIPR measures the extent to which the excitons are delocalized over multiple disks; a dIPR value of $d$ indicates that the exciton has significant amplitude over chromophores on $d$ disks.
\item The middle plot shows the energy and the spatial location of the excitons. The blue line denotes the energy of the exciton. The green line gives the disk number where most of the amplitude of the exciton is localized; i.e. $\textrm{arg max}_n ~p_n^k$ for any exciton indexed by $k$. This quantity constitutes a measure of the approximate spatial location of each exciton. Oscillations in this localization parameter (as a function of $k$) are a heuristic 
\item The bottom plots show the non-zero elements of the modified Redfield rate matrix $R$ describing exciton dynamics in the structure. The magnitude of each element is color coded. These rates are normalized by the largest rate in the matrix and so are all between 0 and 1. This plot nicely illustrates the domains of connectivity in excitonic/energy space. Thus, if two groups of excitons are connected by large matrix elements in $R$, then there will be significant population exchange between them. Individual domains are formed by groups of strongly coupled excitons that are only weakly coupled to other excitons.
\end{enumerate}

In the next section we will refer to these sets of plots in  Figs. \ref{fig:extremes_nodisorder} and \ref{fig:extremes_disorder} to interpret and explain the details of the parametric trends shown in Figs. \ref{fig:parameters_nodisorder} and \ref{fig:parameters_disorder}. We note that for the disordered cases, Figs. \ref{fig:parameters_disorder} and \ref{fig:extremes_disorder}, the parameters, structures, and excitonic details shown are before the introduction of disorder. That is, the physical structures, and consequently excitonic details, shown in these figures are perturbed by each instance of disorder that is sampled over.

\subsection{Optimal structures in the absence of disorder} 

We first compare the optimal values of the 13 design variables for TMV103 and TMV123 without disorder, Fig. \ref{fig:parameters_nodisorder}. We see that the trend in variance of energies of disks is similar for both structures -- i.e. the variance increases steadily as spectral width increases. Thus the incorporation of either distinct chromophores or distinct local energetic environments for these will be necessary to achieve large spectral width in the ordered structures. Furthermore the energy profiles in Figs. \ref{fig:wp_1}(i) and \ref{fig:wp_2}(i) show that the most efficient structures have little heterogeneity in terms of pigment energies, whereas in the opposite extreme, the high-spectral width structures have highly varied pigment energies (although the bottom disk where the trap is situated always has the lowest energy chromophores). The structures in the middle of the Pareto front, i.e., those that achieve a balance between efficiency and spectral width, have a very specific distribution of pigment energies that defines an energy funnel that gradually transitions from high energy pigments at the top to low energy pigments at the bottom of the cylinder. 

Turning to the choice of angular design variables, it is evident that the tangential angle $\theta$ has the most influence on light harvesting function. For TMV103 (Fig. \ref{fig:wp_1}(ii)), the most efficient structures have $\theta \approx 0$ and this angle transitions to a tilted optimal value $\theta \approx 0.6$ as we move along the Pareto front. For TMV123, there is a similar transition from $\theta\approx 0$ for the most efficient structure at the left extremal of the Pareto front, to $\theta \approx 0.4$ for the extremal structure with largest spectral width, although for this more sparse distribution of chromophores, the corresponding transition happens much faster (i.e., at smaller $s$) and the tilted orientation is preferred for most structures on the Pareto front. The radial angle for all structures on the Pareto front for both TMV103 and TMV123 is restricted to be in the range $|\phi | < 0.2$. Therefore having transition dipoles that are tilted into or out of the disk plane does not seem to present any advantage for either efficient transport or spectral width. Within this small range the radial angle can fluctuate and still yield similar values of efficiency and spectral width. Finally, the helical angle $\theta_h$ (whose limits are smaller: $-\pi/17 \leq \theta_h \leq \pi/17$) can also fluctuate significantly without affecting either objective. This suggests that in the absence of disorder, the objectives are fairly robust to both the radial and helical angles (at least in the ranges $|\phi | < 0.2$, $-\pi/17 \leq \theta_h \leq \pi/17$).

To understand the behavior of Pareto structures with respect to the tangential angle, $\theta$, it is useful to also look at the behavior of the ratio of intra-disk and inter-disk dipole couplings $\Lambda$, which is shown in Fig. \ref{fig:parameters_nodisorder}(iii). It is evident that in the absence of disorder, this ratio drops significantly as $s$ increases and that it shows a transition from $\Lambda>1$ to $\Lambda \lesssim 1$ at about the same point as $\theta$ transitions from zero to $\sim 0.6$ ($\sim 0.4$) for TMV103 (TMV123). This transition in $\Lambda$ as a function of $s$ indicates that strong coupling within disks is preferred for improved efficiency, while strong coupling between adjacent disks is preferred for enhanced spectral width.  Consequently the optimal structures show a transition in the direction of the dominant coupling as one moves along the Pareto front.  To confirm that $\theta$ is indeed responsible for this change in the direction of the dominant coupling within the cylindrical array, we plot the parametric dependence of the dipole-dipole coupling between two neighboring pigments on the same disk ($|J_\textrm{\scriptsize intra-disk}|$) and two neighboring pigments on adjacent disks ($|J_\textrm{\scriptsize inter-disk}|$), for TMV103 and TMV123 in Fig. \ref{fig:inter_intra_coupling} . We show here only the parametric dependence on the tangential ($\theta$) and helical ($\theta_h$) angles, since these dependencies are the most relevant. As can be seen from this figure, the coupling between pigments on the same disk dominates for most of parameter space for both TMV103 and TMV123. However, for a region around $\theta\approx 0.6$, this intra-disk coupling becomes very small and is dominated by the inter-disk coupling. This region where the inter-disk coupling dominates is larger for TMV123 because the distance between pigments on the same disk for this structure is larger. So we conclude that the spatial direction of electronic coupling, which is measured by the value of the ratio $\Lambda$, can be controlled by the tangential angle of the transition dipole orientation.

This analysis raises the question as to what is the {\em functional} reason for transitioning from dominant coupling within disks to dominant coupling between disks as we progress along the Pareto front from most efficient to largest spectral width, i.e., as we increase $s$? At first glance, this appears somewhat counter-intuitive, since one might expect that strong coupling between disks would lead to efficient excitation transfer, and conversely that strong excitonic coupling within a disk would lead to spectral broadening. However, the true optimal solutions reflect a more subtle balance of the light harvesting properties possible in this design variable space and geometry. 

Fig. \ref{fig:inter_intra_coupling} shows that in both TMV103 and TMV123, due to the close proximity of pigments within a disk, it is difficult for the coupling between the disks to dominate over the intra-disk coupling. There is only a small region around $\theta \approx 0.6$ where this happens and in this region \textit{both} intra-disk and inter-disk couplings are small ($\lesssim 10$~$\textrm{cm}^{-1}$). In particular, the inter-disk coupling at this angular configuration is dominated by the reorganization energy $\lambda = 100$~$\textrm{cm}^{-1}$ and hence the transport down the cylinder will be F\"orster-like and on slow timescales. Therefore the strategy of simply maximizing coupling between disks to have maximum energy transfer down the cylinder will not work because of the geometric dimensions in these systems. The alternative strategy that is converged on by the optimization of transport efficiency in our calculations, is to have large intra-disk coupling and smaller inter-disk coupling, while maintaining only a slight energy gradient. The slight energy gradient implies that neighboring disks contain almost resonant chromophores (recall that there is no disorder at this stage) and hence even the small coupling between neighboring disks creates excitons that are delocalized across these disks.  We can confirm this strategy by examining the extremely efficient structures in \ref{fig:extremes_nodisorder}. For TMV103 the delocalization of excitons by this strategy is corroborated by the large dIPR values and the oscillation of the disk number in which the excitons have primary spatial locality (both in the left panel of Fig.~\ref{fig:extremes_nodisorder}(a)). This network of moderately delocalized excitons with overlapping spatial locality results in a dense modified Redfield rate matrix (bottom panel of Fig. \ref{fig:extremes_nodisorder}(a)) and efficient population transfer. The most efficient structure in TMV123 utilizes a similar strategy as TMV103, except that the extent of exciton delocalization across disks is a little greater due to the smaller energy gradient (in this case, all disks are almost the same energy). For small values of $s$, TMV123 settles on values of $\theta$ such that the inter-disk and intra-disk couplings are comparable. This creates significant energetic overlap between delocalized excitons that are distant and results in efficient long-range (multi-chromophoric) F\"orster transfer \cite{Jan.New.etal-2004}. These observations are evidenced in the left panel of Fig.~\ref{fig:extremes_nodisorder}(b) that shows the significant delocalization of excitons in the most efficient TMV123 structure, and the dense modified Redfield matrix describing the efficient long-range transfer of energy.

We now consider the structures with largest spectral width, shown in the right panels of Fig.~\ref{fig:extremes_nodisorder}.  Here the strategy for optimizing spectral width in TMV103 and TMV123 is identical. The first element, as already noted is a varied distribution of transition energies: the main mechanism by which spectral width is enhanced is simply by having pigments that absorb at various energies. In addition to this, it is interesting that the angular configurations for the structures maximizing spectral width converge on a coupling ratio of $\Lambda \lesssim 1$. This results in the inter-disk coupling dominating or being the same order as the intra-disk coupling. To understand this, we note that even a small intra-disk coupling is sufficient to create delocalized excitons within disks (because all chromophores on the same disk are resonant), whereas a larger inter-disk coupling is required since chromophores on neighboring disks may be energetically different. Therefore, in order to create a large cluster of coupled chromophores and hence excitons with large delocalization, it is advantageous to maximize the inter-disk coupling while keeping the intra-disk coupling small. This is precisely the strategy utilized in the extremal structures with the largest spectral width, i.e., the large $s$ value structures at the righthand end of the Pareto front. Indeed, the right panels of Fig. \ref{fig:extremes_nodisorder}(a) and \ref{fig:extremes_nodisorder}(b) show that this approach delocalizes excitons over multiple disks (as evidenced by the dIPR and disk population oscillations), which is a signature of electronic coupling of many chromophores over multiple disks. Such electronic coupling of many chromophores results in spectral broadening. It is interesting that the excitons that are most delocalized are the low and high energy ones. This expands the absorption spectrum beyond the limits set by the range of single chromophore transition energies -- i.e. the absorption profiles extend beyond $400$nm and $450$nm. 

\subsection{Optimal structures in the presence of disorder}
\begin{figure}[ht]
\centering
\subfigure[~TMV103 with disorder]{
\includegraphics[scale=0.44]{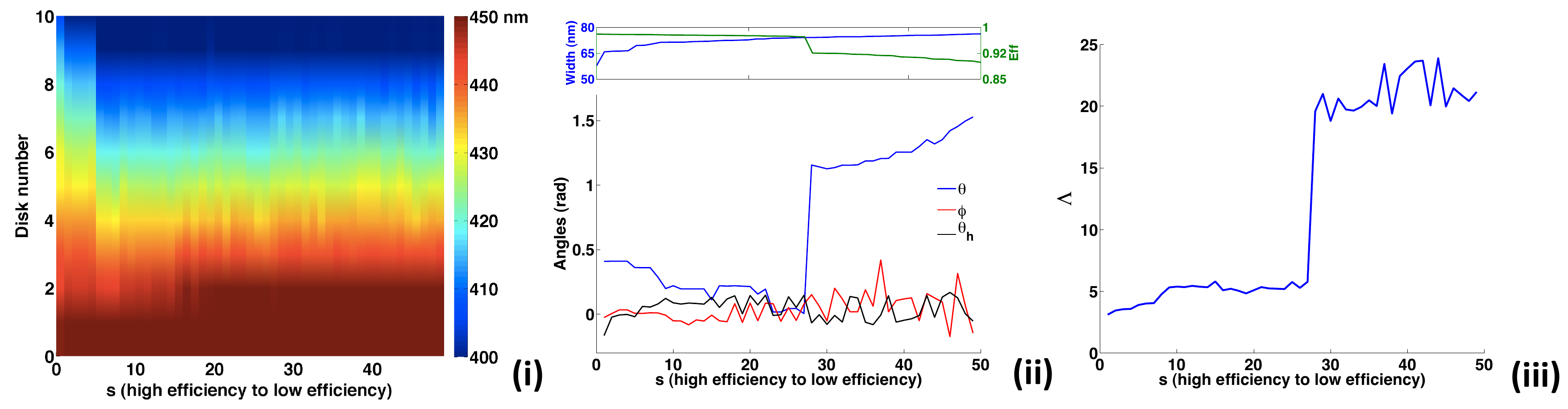}
\label{fig:wp_3}
}
\subfigure[~TMV123 with disorder]{
\includegraphics[scale=0.44]{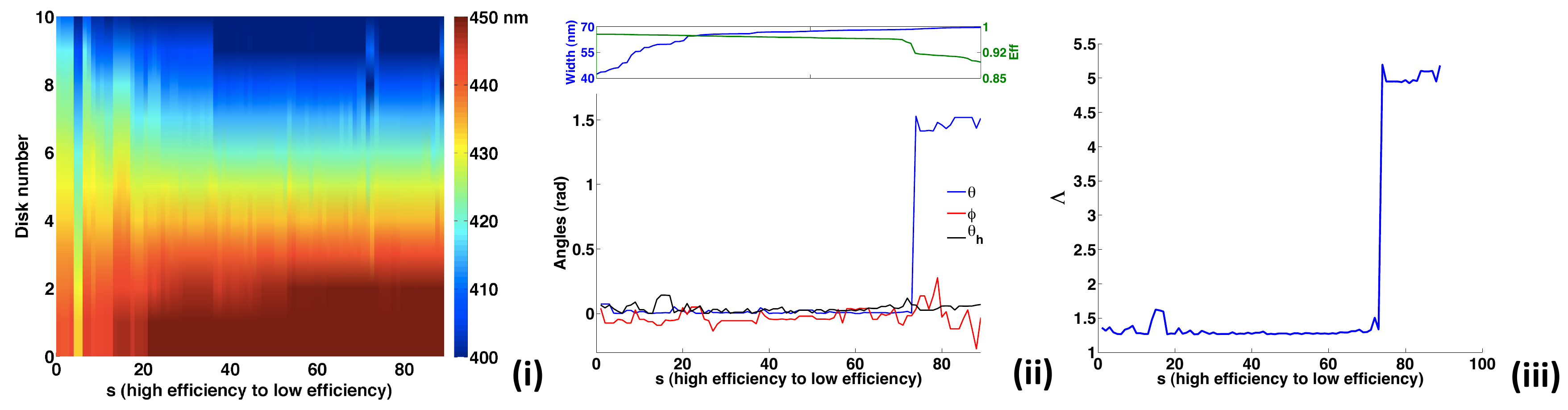}
\label{fig:wp_4}
}
\caption{Design variables for structures on the Pareto fronts \textit{with disorder}, for (a) TMV103 and (b) TMV123. The $x$-axis on all plots shows the structure index $s$ on the Pareto front. Left panels (i): color coded disk energies for the $N=10$ disks. Central panels (ii): transition dipole orientation angles $\theta, \phi$ and helical angle $\theta_h$, with the upper subplot showing the variation of efficiency (green line) and spectral width (blue line) as a function of $s$. Right panels (iii): ratio of intra- to inter-disk dipole-dipole coupling, $\Lambda$. 
\label{fig:parameters_disorder}}
\end{figure}

We now examine the structures and parameters of the optimal configurations on the Pareto fronts when energetic and structural disorder are included. The presence of disorder invalidates many of the optimization strategies outlined 
above for disorder-free systems. Firstly, the most striking observation for the optimal design variables of the disordered systems, is the presence of a strong energy funnel in both Figs. \ref{fig:wp_3}(i) and \ref{fig:wp_4}(i). To achieve any efficiency in the presence of disorder an energy funnel is absolutely necessary, and in both cases (TMV103 and TMV123), a shallow energy funnel amounting to a gradual decrease in pigment energies as the disks approach the bottom layer is more effective at generating efficient transport. With the exception of these most efficient structures, the energetic profiles for the remaining structures on the Pareto front for both TMV103 and TMV123 are remarkably similar. The optimal design strategy for the non-extremal structures appears to be to have discrete decreases in energy every couple of disks, leading to a more step-like energy gradient than the smooth gradient preferred by the most efficient structures. 

\begin{figure}[ht]
\centering
\subfigure[~TMV103 with disorder]{
\includegraphics[scale=0.58]{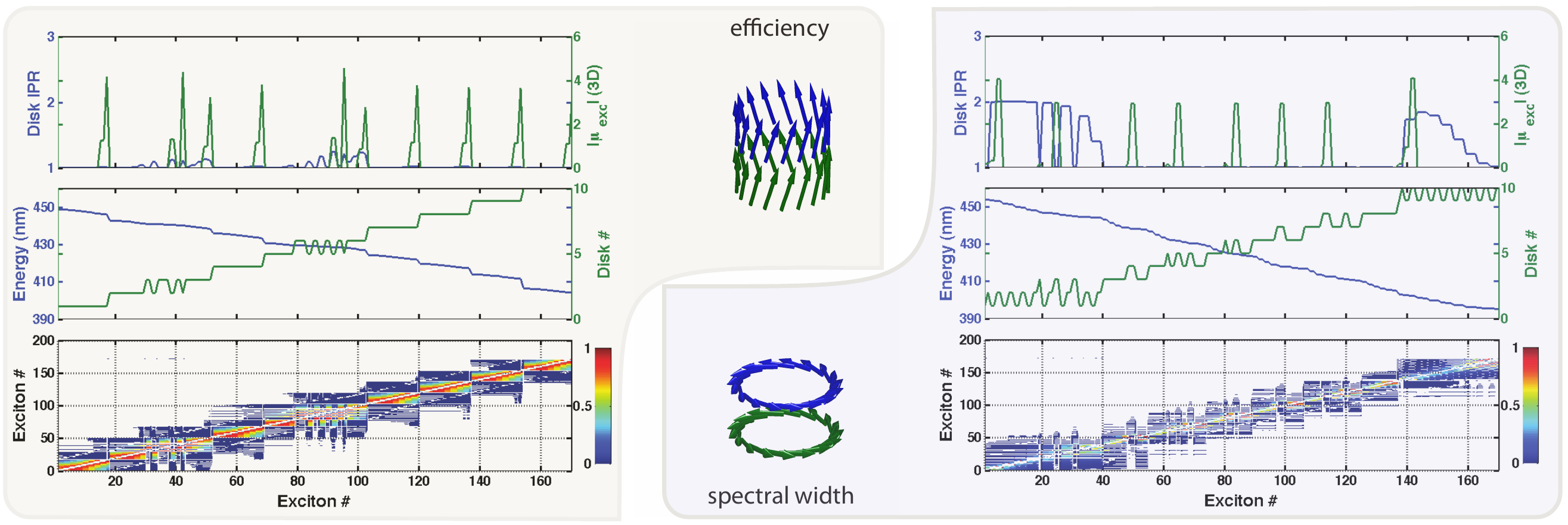}
\label{fig:ex_3}
}
\subfigure[~TMV123 with disorder]{
\includegraphics[scale=0.58]{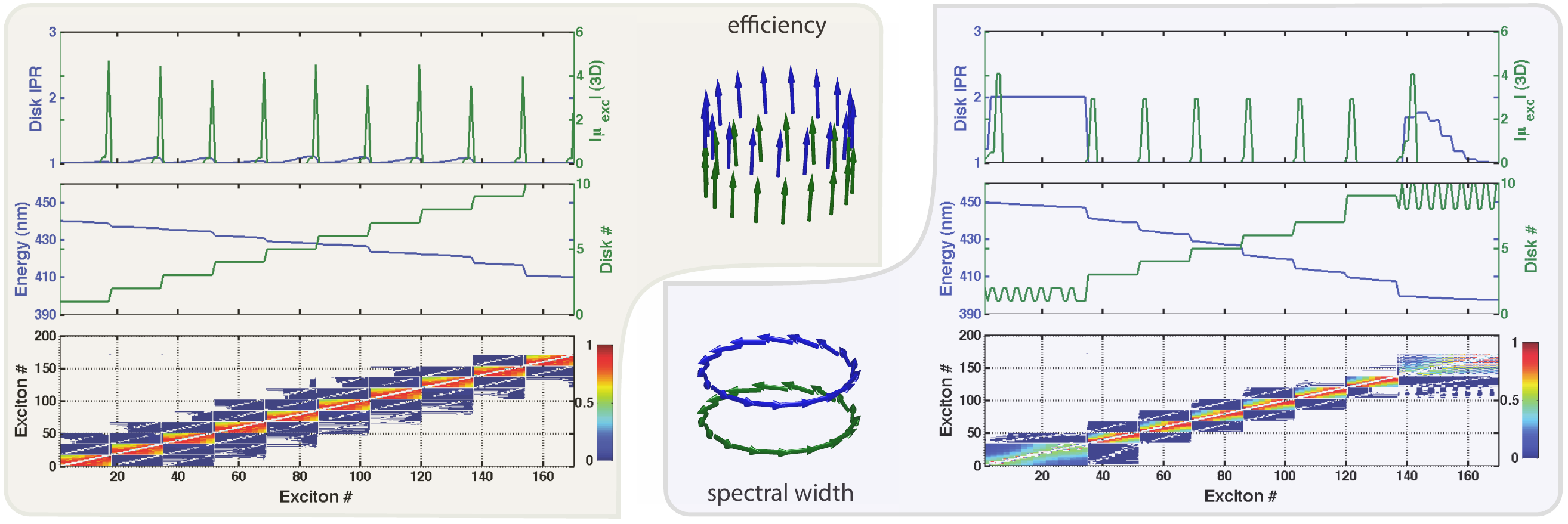}
\label{fig:ex_4}
}
\caption{
Analysis of excitonic structure and dynamics for the extremal structures on the TMV103 and TMV123 Pareto fronts (\textit{with disorder}). See main text (beginning of section \ref{sec:str_params}) for explanation of the plots. 
\label{fig:extremes_disorder}}
\end{figure}

\begin{figure}[h!]
\centering
\subfigure[~TMV103]{
\includegraphics[scale=0.17]{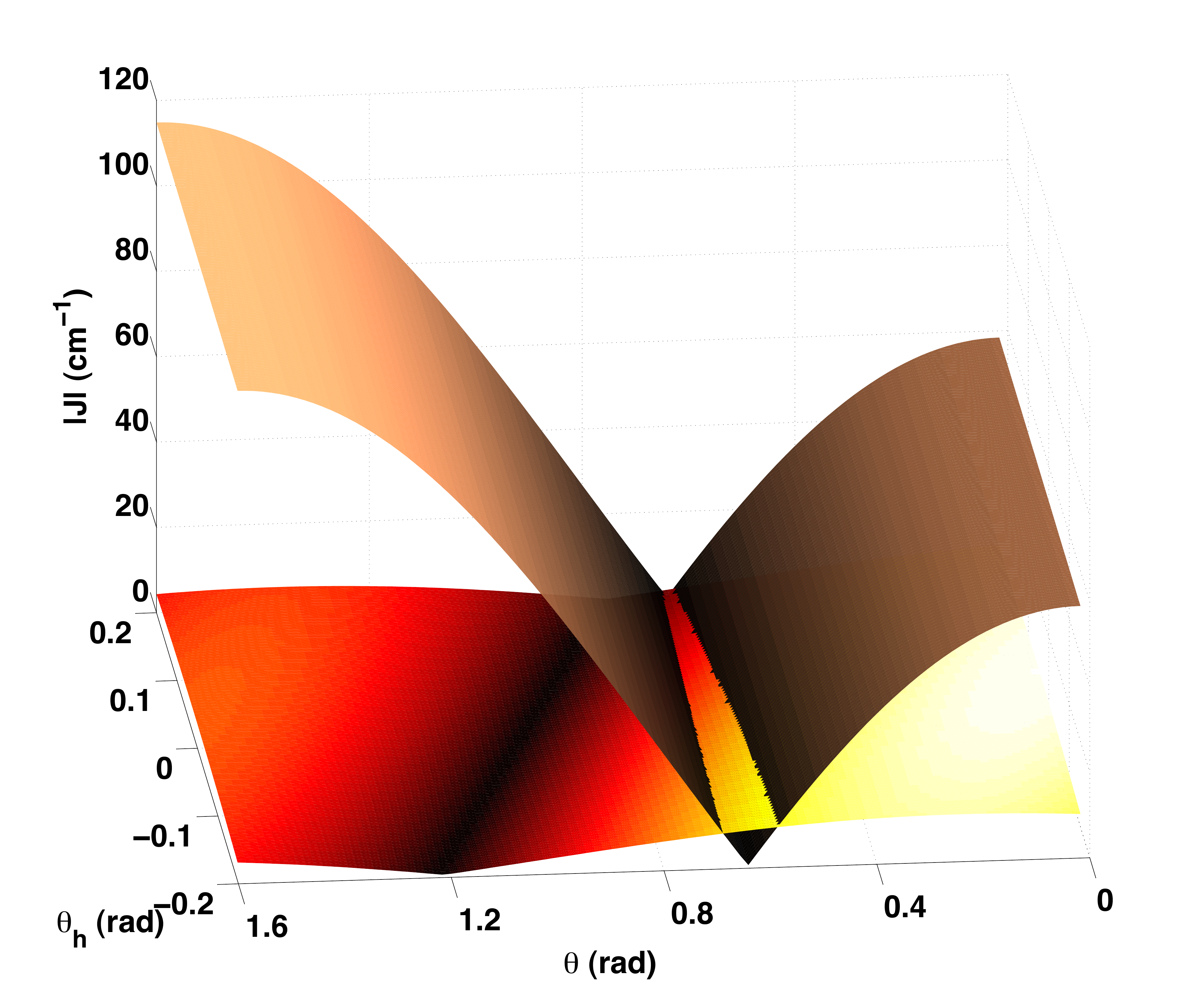}
\label{fig:ii_1}
}
\subfigure[~TMV123]{
\includegraphics[scale=0.17]{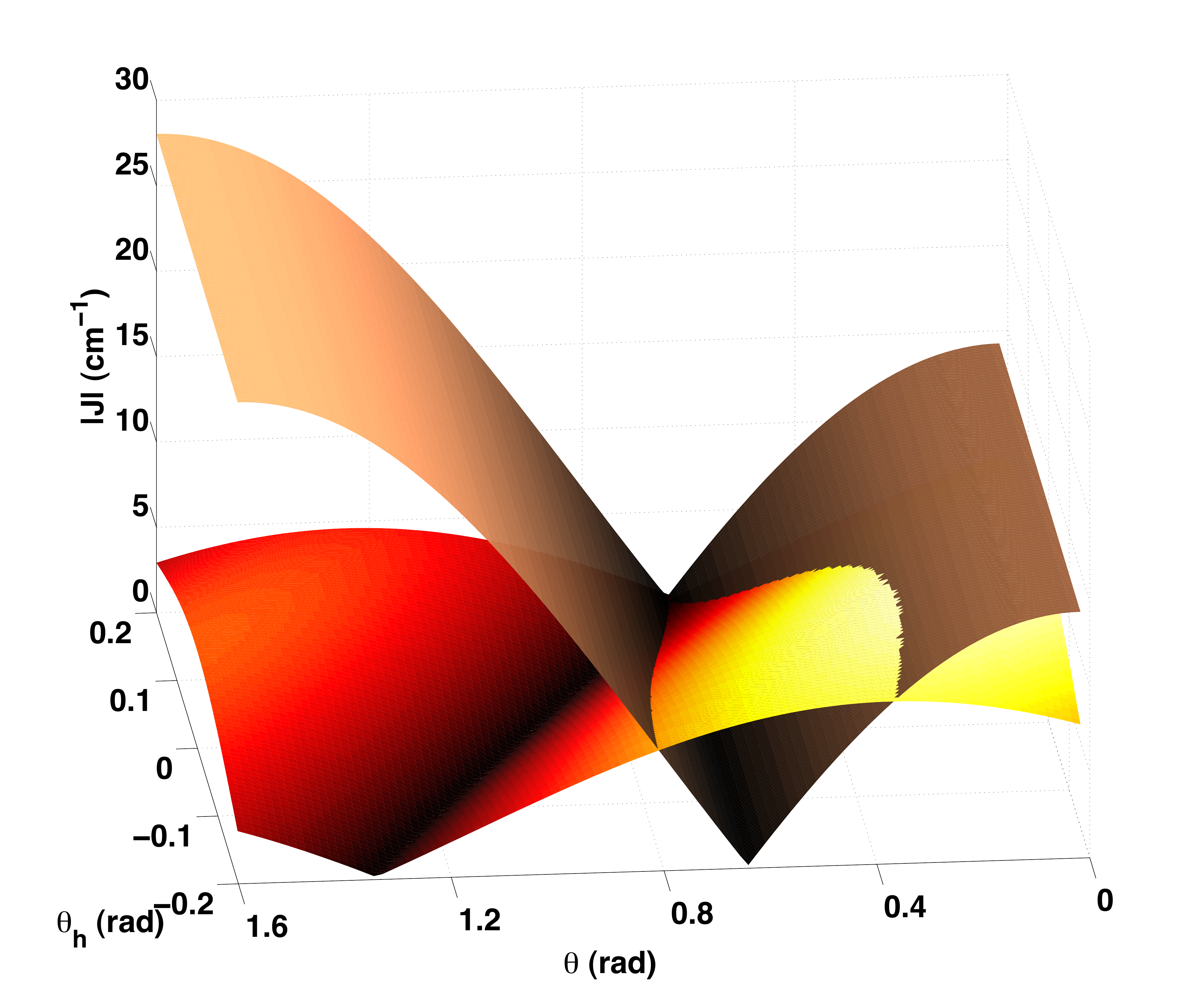}
\label{fig:ii_2}
}
\caption{Dipole-dipole coupling between two neighboring pigments on one disk (brown-tan surface) and two neighboring pigments on adjacent disks (red-yellow surface) as a function of the tangential transition dipole orientation ($\theta$) and the inter-disk helical ($\theta_h$) angles (the radial transition dipole orientation $\phi$ is fixed here at zero). (a) TMV103, (b) TMV123. The primary difference between these two structres is the density of pigments within a disk. The distance between pigments on adjacent disks for both TMV103 and TMV123 is $20\AA$. The distance between pigments on the same disk is $\sim (2\pi /17)\times25 = 9.24$~\AA for TMV103 and $\sim (2\pi /17)\times40= 14.78$~\AA for TMV123.  \label{fig:inter_intra_coupling}}
\end{figure}

Turning to the choice of angular design parameters, consider the case of TMV123 first (Fig. \ref{fig:wp_4}(ii)).
Unlike the disorder-free case, where a variety of angle choices (especially for $\phi$ and $\theta_h$) yielded similar objectives, in the presence of disorder a few optimal configurations emerge. In particular, the most efficient structures have $\theta = \theta_h \approx 0$. At the other extreme, the structures with the largest spectral width have $\theta \approx \pi/2$ and $\theta_h \approx 0$. The radial angle is constrained to be $| \phi | < 0.25$ and is close to zero for nearly all structures on the Pareto front. As with the disorder-free case, we find that the tangential angle has the biggest influence on the objectives. The radial and helical angles have only a slight influence, as long as they are constrained to be $\approx 0$. For disordered TMV103 (Fig. \ref{fig:wp_3}(ii)), the tangential angle again has the biggest influence and it transitions from $\theta \sim 0.4$ to $\theta \sim \pi/2$ as $s$ increases. The other angles fluctuate along the Pareto front of TMV103 more than for TMV123, but we cannot discern any pattern in their variation along the Pareto front. Finally, Figs. \ref{fig:wp_3}(iii) and \ref{fig:wp_4}(iii) show that there is a clear preference for having dominant coupling along the cylinder (low $\Lambda$) to optimize efficiency, while dominant coupling within disks is preferred in order to optimize spectral width. In fact the most efficient structures {\em maximize} the inter-disk coupling. For TMV103 this is achieved around $\theta \approx 0.4, \theta_h=0$ and for TMV123 it is achieved around $\theta \approx 0, \theta_h \approx 0$ (see Fig. \ref{fig:inter_intra_coupling}). And similarly, the structures with maximum spectral with \textit{maximize} the intra-disk coupling, which occurs at $\theta \approx \pi/2$ as seen from Fig. \ref{fig:inter_intra_coupling}.

We can gain more insight into the parameter landscape by examining the extreme structures on the disorder Pareto fronts. As the analysis below will show, the inability to construct large delocalized excitonic states on any timescale in the presence of significant static disorder dominates the design choices made in optimizing light harvesting. The left panels of Figs. \ref{fig:ex_3} and \ref{fig:ex_4} show that the most efficient structures for TMV103 and TMV123 have similar excitonic structure, showing several distinct features. Firstly, the excitons are mostly localized on a single disk. There is a slight delocalization, especially in TMV103, as evidenced by the dIPR and oscillations in disk population but it is minimal compared to the disorder-free structures. This is a result of the strict energy gradient, with each disk having pigments with a different transition energy and consequently little delocalization across disks. The addition of static disorder, and also dynamic disorder induced by protein fluctuations, will of course further localize the excitons. The structure of the modified Redfield rate matrix for the most efficient TMV103 and TMV123 structures (bottom plots of left panels Fig. \ref{fig:extremes_disorder}) is seen to be fairly sparse with small rates between densely coupled domains. The domains are localized on disks and there is larger transfer rates between excitons within a domain than between domains. This localization of excitons into domains and inter-domain dominated transport is reminiscent of multi-chromophoric F\"orster transfer \cite{Jan.New.etal-2004}, and we can expect this to provide accurate description of the energy transfer in these structures optimized for efficiency in the presence of disorder. The strategy here will be to have slightly delocalized excitons on disks that are as strongly coupled as possible to excitons on neighboring disks due to maximized inter-disk electronic coupling (at $\theta \approx 0.4, \theta_h=0$ for TMV103 and at $\theta \approx 0, \theta_h \approx 0$ for TMV123), and a shallow energy gradient which ensures that neighboring disks contain chromophores that are nearly resonant despite the disorder. 

Now we turn to the extremal structures that maximize spectral width in the presence of disorder. These are illustrated with the lower structures shown in the central panels of Fig. \ref{fig:extremes_disorder}.  There is little obvious difference between these two extremal structures for TMV103 and for TMV123.  Both consist of disks that are J-aggregate-like, with dipoles aligned head-to-tail. The energy of each disk is different, except for the first two and last two which have almost identical energies. This creates a series of almost independent J-aggregates (i.e. bright excitons are the lowest energy ones) that absorb at a range of energies \footnote{Note that the excitonic details shown in Fig. \ref{fig:extremes_disorder} are for the ideal design parameters. Each instance of disorder will perturb these and result in a perturbed version of these excitonic details. In particular, the exciton delocalization for the maximum spectral width structures (for TMV103 and TMV123) seems large from the dIPR values shown in the right panels of  Fig. \ref{fig:extremes_disorder}. However, any instance of disorder will break the symmetries of the ideal design and result in more localized excitons.}.  The reason for this form of the optimal structure for achieving maximal spectral width can be understood by examining Fig. \ref{fig:inter_intra_coupling}, which reveals that the maximum possible excitonic coupling between any two chromophores (for TMV103 and TMV123) is an intra-disk coupling that results from $\theta = \pi/2$, i.e., from dipoles aligned head-to-tail within a disk. This strong coupling is advantageous for competing against disorder and maintaining excitonic coupling in its presence. It is not, however, advantageous for efficient transport down the cylinder since the strong coupling is intra-disk rather than inter-disk, but it is nevertheless advantageous for spectral broadening. 
We can rationalize this finding with a simple model for the effect of disorder on coupling of any two chromophores.  Thus, given excited states of two chromophores with energies $E_1, E_2$ and electronic coupling $J$, the resulting excitonic states have energies $$e_{\pm} = \frac{E_1+E_2}{2} \pm \frac{\sqrt{(E_1-E_2)^2 + 4J^2}}{2}.$$ Assuming that neither are dark states (\textit{i.e.} have zero net dipole strength), the absorption peaks of the coupled system will be centered around $e_\pm$. In presence of finite coupling $J$, these energies can be different from $E_1$ and $E_2$, causing spectral broadening due to excitonic coupling. In the presence of disorder, $|E_1-E_2|$ can be large, in which case $e_\pm$ will only be significantly different from the original transition energies ($E_1$ and $E_2$) if $J$ is comparably large. This is the reason for the J-aggregation seen in the maximum spectral width structures; this mechanism achieves the strongest coupling and thus enables broadening even in the presence of disorder. The spectral widths achievable with disorder for TMV103 are slightly larger than those achievable for TMV123 because the former system can attain greater intra-disk coupling strengths due to the greater proximity of pigments within a given disk (see Fig.~\ref{fig:inter_intra_coupling}).  In summary, in the presence of disorder, maximal spectral width is achieved by having each disk effectively act as its own antenna with strong intra-disk coupling present to provide spectral broadening and with almost every disk composed of pigments with distinct transition energies. Interestingly, this optimized structure bears resemblance to multi-junction or tandem solar cells \cite{Cotal:2009dk, Ameri:2009jv}, which are also designed to increase the range of wavelengths of utilized photons.

It is interesting to note that the strategies found here for optimizing spectral width with and without disorder are only effective in their respective cases. That is, the disorder strategy of creating J-aggregate-like maximally coupled disks would not be effective in the absence of disorder. This is because of the well known property of J-aggregates that they possess red-shifted, narrow lines of absorption because the structural symmetry renders most states optically inaccessible (\textit{i.e.} results in dark states). Therefore in the absence of disorder the absorption profile would consist of thermally broadened peaks around the J-aggregate absorption lines, which is unlikely to produce a very wide spectrum. However in the presence of disorder, the complete $J$-aggregate symmetry is broken and the majority of states are no longer dark states, allowing wide bandwidth absorption and effective spectral broadening at the same time (note that the maximum disorder should be comparable in magnitude to the maximum achievable electronic coupling for this strategy to be effective). Disorder is thus essential to the success of this strategy. Similarly, the disorder-free strategy that is optimal for maximizing spectral width is to use the spectral broadening resulting from excitonic coupling across many chromophores. This strategy is in turn ineffective in the presence of disorder, since this would suppress such couplings and resulting broadening, as seen from the above equation for a two chromophore system. These observations underscore the important point that the optimal strategies for light harvesting can change dramatically depending on the amount of disorder present in the system. 

Finally, we comment on the effects of changing the non-Hamiltonian components of the system. Although we have not optimized over the uncontrollable degrees of freedom, we have performed multi-objective optimizations for several values of the reorganization energy and several degrees of disorder. The main functional influence of the reorganization energy is to vary the amount of absorption linewidth broadening. For example, when the reorganization energy was doubled to $\lambda=200\textrm{cm}^{-1}$ the achievable absorption linewidths for TMV103 and TMV123 increased. However, the shapes of the Pareto fronts remained the same and the relative advantage that TMV103 has over TMV123 in the presence of disorder was preserved (i.e. TMV103 is able to achieve greater absorption widths while maintaining efficiency than TMV123). The optimal structures also remained similar, with the same patterns of change in design variables when moving across the Pareto fronts. In contrast to changes in reorganization energy, changes in the amount of disorder were seen here to dramatically effect the optimal structures. This is already evident when comparing the Pareto fronts and optimal structures in cases of zero disorder and finite disorder presented above. We find that when the structural and energetic disorder are increased further they can dominate the light harvesting function, with structural optimizations playing less of a role. In such cases, the formation of an energetic gradient is the most effective design principle. We consider the situation that we have presented above to be the most interesting since with this choice of disorder, the fluctuations in electronic coupling energies due the disorder are comparable to the other energy scales in the system (e.g. thermal energy, reorganization energy). This results in many design parameters playing an active role in light harvesting performance and in the most interesting optimization structure. For future work, it would be interesting to incorporate experimentally determined disorder parameters, as they become available, into these multi-objective optimization studies. 

\section{Design principles for engineering light harvesting antennas}
\label{sec:design}
Given the understanding we have gained of the structures on the Pareto front and their relative performance, we can now extract several design principles for optimal light harvesting using cylindrical chromophore assemblies. 

\begin{enumerate}
\item A trade-off between efficiency and spectral width exists for cylindrical light harvesting antennas. There are limits to maximizing both of these objectives, especially in the presence of disorder. 
\item Optimal structures for efficiency or spectral width change significantly depending on whether energetic and structural disorder is present or not. In the absence of disorder, many choices of the design variables yield similar values of the objective functions. However, when disorder is present only a few choices of the design variables are truly optimal -- i.e., the design variables for neighboring structures on the Pareto front are fairly similar. 
\item An energy funnel is essential regardless of the density of pigments for such structures with a single dedicated sink for excitons. We see the emergence of an energy funnel for both TMV103 and TMV123. A shallow energy funnel is the most effective for transport. 
\item The attachment angle that dictates the tangential angle of the dominant dipole ($\theta$) is the most critical orientation degree of control, followed by the helical angle $\theta_h$. The radial angle $\phi$ should be kept as close to zero as possible since all structures on the Pareto front retain this property. 
\item An increased density of pigments does not aid in increasing efficiency or spectral width in the absence of disorder. However, in the presence of disorder, increased density is advantageous since it can lead to stronger electronic couplings which can be used to combat the deleterious effects of disorder. 
\item Simply having access to strong electronic coupling is not sufficient to overcome the deleterious effects of disorder. The direction of this coupling has to be controlled and used effectively. In the case of TMV103 and TMV123 the strongest couplings are intra-disk simply because of the dimensions. These couplings only help to maintain spectral width in the presence of disorder and do not directly generate efficient energy transfer (although other strategies help to generate efficient transfer).
\end{enumerate}

We note that in the present system with a full complement of $M=17$ chromophores per disk, there does not appear to be any apparent benefit from having a helical twist to the chroomophore arrangement along the cylinder.  This may simply reflect the relatively close packing of chromophores in this system and the ability of the other 12 parameters to simultaneously optimize the energy efficiency and spectral width.  However it may also reflect the lack of any chiral constraint element at the chromophore level.  More extensive calculations with smaller values of $M$ and more refined models of chromophores are needed in order to determine whether additional structural constraints such as a center of chirality on the chromophores are necessary for achieving helical optimal structures for these design objectives.  

We can now ask how many of these design principles are manifest in natural photosynthetic light harvesting antennae. Such observations are unavoidably speculative since we do not understand the structure-function relationships in natural antennae sufficiently well at this stage to make general statements. Nevertheless, it is still interesting to compare these emergent design principles to features and motifs seen in natural systems. Firstly, consider chlorosomes, the main light harvesting antennas of green sulfur bacteria, which are cylindrical antennas consisting of very densely packed BChl \textit{c, d}, or \textit{e} molecules \cite{Prokhorenko:2000uc, Gan.Oos.etal-2009}. Given the above results, it is plausible that the dense packing found in chlorosomes has the primary function of increasing the range of usable photons, an important feature for green sulfur bacteria, which typically live in severely energy limited environments. It is difficult to speculate on the impact of structure on energy transfer efficiency because it is unknown at present how the photo-excitation couples out of the chlorosome complex and hence we cannot model energy extraction simply as trapping sites at the base of the cylinder as done above. One could also speculate that the small energy gradient seen in PS-II \cite{Jennings:1996ut} is consistent with the above observation above that small energy gradients are beneficial for maximum efficiency of transport in disordered systems.

\section{Conclusion}
\label{sec:conclusions}
Inspired by the cylindrical assemblies of chromophores that can be synthesized by TMV-templated assembly, we have examined the landscape of light harvesting performance achievable by such structures. In particular, we have studied the trade-offs involved in optimizing over multiple objectives relevant to light harvesting. In addition to identifying a fundamental trade-off between optimizing energy transfer efficiency and bandwidth of absorption, our calculations have allowed the development of several design principles to guide the design and construction of such cylindrical assemblies of chromophores. We find that the presence of disorder drastically effects the optimal structural forms and therefore it is essential to characterize the amount of structural and energetic disorder present in typical TMV-templated chromophore assemblies. Experiments to-date on protein-templated systems are insufficient to characterize this degree of disorder, and hence this is a critical task for future experiments. The design principles established here provide guidelines for a program of quantum-informed rational design (QuIRD) for biomimetic light harvesting systems.

Our study also lends insight into the structure of biological photosynthetic light harvesting complexes. The multi-objective optimization study clarifies the potential role of strong electronic couplings enabled by high densities of chromophores. In realistic systems with disorder, our study suggests that strong electronic couplings are somewhat beneficial for maintaining efficiency of energy transfer, but more importantly, they are essential for increasing the spectral width of absorption profiles. Therefore the strong electronic couplings recently observed in several biological LHCs -- e.g. Refs. \cite{Eng.Cal.etal-2007, Col.Won.etal-2010, Pan.Hay.etal-2010, SchlauCohen:2012dn} -- may not only aid excitation transport in such systems, but also simultaneously benefit optical measures of light harvesting performance such as spectral width of absorption. 

This first multi-objective optimization of structural design for biomimetic light harvesting systems suggests several interesting directions for further work. The first avenue for future work is the consideration of measures of performance other than absorption bandwidth. Although efficiency of energy transfer is likely to be desirable in all applications, other measures of performance in the optical domain could be considered. For example, sensitivity to photons of a particular wavelength could be relevant for the design of sensor technologies. As another avenue of future work, we plan to carry out a multi-objective optimization study of the helical structural that is templated using TMV \cite{Klu-1999}. The helical assembly could be more stable than the stacked disk assembly considered in this work, however features such as an energy gradient could be more difficult to implement in the helical structures. 

It would also be interesting to simulate excitation dynamics for the optimal structures on the Pareto front using a non-perturbative technique -- such as the hierarchical equations of motion formalism \cite{Ish.Fle-2009} -- that captures dynamical coherence (time-dependent coherence between excitons). Although the spectral characteristics will be unaffected by such treatments, they will refine the measure of energy transfer efficiency and also give an explicit indication for how important dynamical coherence is to the effectiveness of energy transfer in such structures. Unfortunately, the TMV-templated chromophore assemblies are large (in terms of number of chromophores), and beyond the limit of what is currently computationally feasible using such non-perturbative techniques. However, this direction could become feasible in the future. 

Finally, we plan to perform multi-optimization studies of TMV-templated chromophore assemblies that are restricted to varying the experimental degrees of freedom that are currently tunable. That is, the design variables will be restricted to the ones that are currently experimentally accessible, so that the optimized designs could be constructed and verified in the immediate future.

\section{Acknowledgements}
We gratefully acknowledge Matthew Francis and Daniel Finley for useful discussions relating to virus-templated chromophore assemblies.  We thank Akihito Ishizaki for bringing to our attention the continued fraction solution in Ref. \cite{Takagahara:1977wq}. Financial support for MS and KBW was provided by the DARPA QuBE (Quantum Effects in Biological Environments) program. The views expressed are those of the authors and do not reflect the official policy or position of the Department of Defense or the U.S. Government. Some of the multi-objective optimization were performed using supercomputer time allocated through XSEDE. Sandia National Laboratories is a multi-program laboratory managed and operated  by Sandia Corporation, a wholly owned subsidiary of Lockheed Martin Corporation, for the United States Department of Energy's National Nuclear Security Administration under contract DE-AC04-94AL85000.

\section*{References}
\bibliographystyle{unsrt}
\bibliography{lhc_tradeoffs}

\newpage

\appendix
\section{Multi-objective optimization}
\label{sec:mo_opt}
The optimization of multiple objectives is a commonly encountered problem in engineering and sciences. Given a set of control variables, $\vec{x} \equiv (x_1, x_2, ..., x_N)$, the task of multi-objective optimization, in this example a maximization, is:
\beq
\max_{\vec{x}} ~ [ f_1(\vec{x}), f_2(\vec{x}), ... f_M(\vec{x}) ]
\eeq
where $f_i(\vec{x})$ are $M$ objectives (cost functions). One could, and usually does, also have equality or inequality constraints on the variables. For convenience we notate $\vec{f} \equiv  [ f_1(\vec{x}), f_2(\vec{x}), ... f_M(\vec{x}) ]$. This optimization over multiple objectives has multiple non-degenerate solutions unlike a single objective optimization which has a single solution (or multiple degenerate ones). The set of solutions are referred to as \textit{Pareto points} and are the set of points in objective space such that improving any one objective can only be done at the expense of another. Specifically, $\vec{f}$ is a Pareto point, or is Pareto optimal, if there does not exist another feasible objective vector $\vec{f}'$ such that $f_i' \geq f_i ~~ \forall i\in \{1,2,..., M\}$, and $f_j' > f_j$ for at least one $j\in \{1,2,..., M\}$. 

The set of Pareto points in objective space is also called the \textit{Pareto front}. For each Pareto point in objective space, $\vec{f}^*$, there is a corresponding Pareto point in control variable space, $\vec{x}^*$ and is specified by the control variables that achieve $\vec{f}^*$. 

One approach for solving multi-objective optimizations is to combine the multiple objectives into one objective, e.g. $F(\vec{x}) = \sum_{i=1}^M \alpha_i f_i(\vec{x})$ with $\alpha \in \mathbb{R}$ is a linear combination of objectives. Then a conventional single objective optimization solution method is used to solve this problem. This strategy is acceptable when the acceptable combination of objectives is known \textit{a priori}. However, in many problems one cannot know ahead of time how to combine the multiple objectives into one. In addition, knowledge of the Pareto front for the full multi-objective problem is valuable for understanding the types of trade-offs involved in the optimization problem. In this work, where we are particularly interested in characterizing the trade-offs involved in light harvesting, the full multi-objective optimization ins critical. 

There are a variety of methods for solving multi-objective optimizations \cite{Marler:2004ha}. In this work we employ an evolutionary algorithm which has the advantage that it requires few assumptions and \textit{a priori} knowledge of the optimization landscape. We utilize the non-dominated sorting genetic algorithm II (NGSA-II) by Deb \textit{et al.} \cite{Deb:2002ut} implemented in the C++ optimization toolbox written by Sastry \cite{Sastry:2007ww}. The specific genetic algorithm parameters we use are:
\begin{enumerate}
\item Population size: 100
\item Number of generations evolved: 100
\item Proportion of population replaced in each generation: 0.7
\item Crossover probability (with simulated binary crossover): 0.85
\item Mutation rate (with polynomial mutation): 0.3
\end{enumerate}
These parameter values were converged upon by experimentation and determining what combination of population size, mutation rate and crossover rate allowed good exploration of the optimization landscape. For each optimization, multiple runs were executed (there were several restarts), each from a different randomly chosen starting population and all the resulting populations were collected. This distribution of starting points added another degree of randomization to ensure that the large optimization landscape was sampled reasonably well. 

\section{Modified Redfield theory}
\label{sec:dyn_model}
The equations of modified Redfield theory specify rates of exciton population transfer, and define a rate equation for the exciton population dynamics:
\beq
\frac{\textrm{d}P(t)}{\textrm{d}t} = R P(t)
\label{eq:rate_eqn}
\eeq
where $R$ is the modified Redfield rate matrix \cite{Yan.Fle-2002}, and $P(t)$ is a vector of exciton populations. In this work we only consider exciton dynamics in the single exciton manifold since this is the most relevant at low to moderate solar irradiance. As derived in Ref. \cite{Yan.Fle-2002}, the rate for population transfer from exciton $k$ to $k'$ is:
\beq
R_{k,k'} = 2 \textrm{Re}\int_0^\infty d\tau F^*_{k'}(\tau)A_k(\tau) N_{k,k'}(\tau)
\eeq
with

\bqa
F_{k'}(\tau) &=& \exp(-i(E^0_{k'}-\lambda_{k'})\tau - g^*_{k'k',k'k'}(\tau)) \nn \\
A_k(\tau) &=& \exp(-i(E^0_{k}+\lambda_{k})\tau - g_{kk,kk}(\tau)) \nn \\
N_{k,k'}(\tau) &=& \bigg(\ddot{g}_{k'k,kk'}(\tau) - [\dot{g}_{k'k,kk}(\tau) - \dot{g}_{k'k,k'k'}(\tau)-2i\lambda_{k'k,k'k'}] \nn \\
&& \times [\dot{g}_{kk',kk}(\tau) - \dot{g}_{kk',k'k'}(\tau)-2i\lambda_{kk',k'k'}] \bigg)~ e^{2(g_{kk,k'k'}(\tau)+i\lambda_{kk,k'k}\tau)} \nn
\eqa
Here, $E^0_k=E_k - \lambda_k$ is the $0-0$ exciton transition energy -- i.e. $E_k$ is the exciton transition energy and $\lambda_k$ is the exciton reorganization energy, defined by $\lambda_k \equiv \sum_{n=1}^K  |U_{n,k}|^4 \lambda_n$ with $U_{n,k}$ the exciton-site basis transfer coefficients and $\lambda_n$ the reorganization energy of site $n$. Similarly, $\lambda_{\alpha\beta,\gamma\delta} \equiv \sum_{n=1}^K U^*_{n,\alpha}U_{n,\beta}U^*_{n,\gamma}U_{n,\delta} \lambda_n$, and the exciton lineshape function is defined as $g_{\alpha\beta,\gamma\delta}(t) \equiv \sum_{n=1}^K U^*_{n,\alpha}U_{n,\beta}U^*_{n,\gamma}U_{n,\delta} g_n(t)$ where $g_n(t)$ is the single chromophore (site) phonon-induced lineshape function. In this work we assume all chromophores have the same lineshape function, and it derives from a high temperature over-damped Brownian oscillator model of the phonons with spectral density $J(\omega) = \frac{2\lambda \gamma \omega}{\omega^2 + \gamma^2}$. In this case,
\beq
g(t) = \left( \frac{2\lambda}{\beta \hbar^2 \gamma^2}-\frac{i\lambda}{\hbar \gamma}\right)\left(e^{-\gamma t}-1+\gamma t\right)
\eeq

The integrals defining the modified Redfield rates are time-consuming to perform numerically for an arbitrary lineshape function, but in the case of the model above, the integral defining the modified Redfield rates becomes of the form:
\beq
R_{k,k'} ~\propto~ 2\textrm{Re} \int_0^\infty d\tau ~e^{c_1 \tau - c_2 e^{-\gamma \tau}}
\eeq
where $c_i$ are $k,k'$ dependent, but time-independent, complex coefficients. Integrals of this form can be performed analytically using a continued fraction expansion, as shown by Takagahara \etal in the Appendix of Ref. \cite{Takagahara:1977wq}. We use this continued fraction solution to calculate the modified Redfield rates efficiently. 

Now we detail how to extract a measure of spectral width of absorption and efficiency of transport from the modified Redfield model of exciton dynamics.

\subsubsection{Spectral width}
An expression for the absorption spectrum in terms of the modified Redfield exciton transfer rates is \cite{Nov.Dou.etal-2010}:
\beq
A(\omega) = \omega \sum_k \mathbf{d}_{k}^2 \textrm{Re}\bigg\{ \int_0^\infty \textrm{d}t \exp \bigg( i(\omega - \omega_k)t - \sum_{n=1}^K |U_{n,k}|^4 g(t) - \frac{t\gamma_k}{2} \bigg) \bigg\}
\eeq
where the $k$ sum is over all excitons. $\mathbf{d}_k$ and $\omega_k\equiv E_k/\hbar$ are the magnitude of the transition dipole and the transition frequency  of exciton $k$, respectively. $g(t)$, as detailed above, is the phonon induced lineshape function, taken to be that of an over-damped Brownian oscillator and the same for all chromophores. Finally, $\gamma_k$ is the inverse lifetime of exciton $k$, given by a sum of outgoing modified Redfield rates: $\gamma_k = -\sum_{k' \neq k} R_{k'k'kk}$.  

We construct the absorption spectrum according to this expression and normalize it to have maximum 1. Then the width of the absorption spectrum is defined as the sum of frequency intervals for which $A(\omega) > 0.1$.

\subsubsection{Efficiency}
Consider the modified Redfield rate equation, \erf{eq:rate_eqn}, which describes the dynamics of the exciton populations. In this work, we use $P(t) \equiv (p_1(t), p_2(t), ..., p_K(t), p_\textrm{trap}(t), p_\textrm{loss}(t) )^T$, where the first $K$ elements are the populations of the $K$ excitons at time $t$ (in the main text we consider cylindrical TMV assemblies with $N$ disks and $M$ chromophores per disk, and therefore $K=N\times M$). The remaining two elements in this population vector are the population in the trap (i.e. the fraction of excitons that have undergone charge separation), $p_\textrm{trap}(t)$, and the population lost to recombination, $p_\textrm{loss}(t)$. The elements of $R$ that represent rates of transfer between excitonic populations are prescribed by modified Redfield theory as above, and the rates of transfer from exciton $k$ to the trap ($R_{k\rightarrow \textrm{trap}}$) and loss ($R_{k\rightarrow \textrm{loss}}$) channels are defined as:
\bqa
R_{k \rightarrow \textrm{trap}} &= \gamma_\textrm{trap}\sum_{n \in \textrm{disk N}} |U_{n,k}|^2 \nn \\
R_{k \rightarrow \textrm{loss}} &= \gamma_\textrm{loss} \mathbf{d}_k^2
\eqa
The motivation behind that $k \rightarrow \textrm{trap}$ rate is that exciton $k$ has probability $\sum_{n \in \textrm{disk N}} |U_{n,k}|^2$ of being localized at one of the chromophores on the bottom most disk (disk $N$) where the trapping (charge separation) is assumed to take place. Hence the total trapping rate is this probability multiplied by the bare trapping rate, $\gamma_\textrm{trap}$. In this work we choose $1/\gamma_\textrm{trap} = 4$ps. The motivation behind the $k \rightarrow \textrm{loss}$ rates is that the rate of radiative recombination of an exciton is equal to a bare rate of recombination $\gamma_\textrm{loss}$ multiplied by the dipole strength of the exciton. This ensures that bright and dark excitons recombine at different rates. In this work we choose the bare rate of recombination as $1/\gamma_\textrm{loss} = 1$ns.

We define the efficiency of transport as the asymptotic fraction of population in the trap, $\eta = p_\textrm{trap}(t\rightarrow \infty)$. Using a Laplace transform solution of the rate equation \erf{eq:rate_eqn} results in the following expression for this asymptotic population:
\beq
P(t\rightarrow \infty) = \lim_{s\rightarrow 0} (s\mathbf{I}-R)^{-1}P(0)
\eeq
where $P(0)$ is the initial population distribution. We approximate this asymptotic solution by taking a small value of $s$ in the above expression. We have confirmed that our choice of $s$ does not influence the solution, especially the value of $\eta = p_\textrm{trap}(t\rightarrow \infty)$.

Our choice of $P(0)$, the initial exciton distribution is dictated by the size of each exciton's dipole. That is, $p_k(0) = \mathbf{d}_k^2/\mathcal{N}$, where $\mathcal{N}$ is a normalized so that $\sum_{k=1}^K p_k(0) =1$. This initial state specifies that the probability that an exciton is initially populated is proportional to its transition dipole strength. The initial trap and loss populations are set to zero. 

Finally, we comment on alternatives to asymptotic efficiency for measures of energy transfer effectiveness. The average trapping time, defined as the time at which the trap population is greater than $1-\epsilon$ for some small $\epsilon$, is a measure that captures the speed at which the excitation is transferred. This measure could be more useful than asymptotic efficiency when further stages of the light harvesting process (e.g. charge separation) are modeled and optimized. This is because for such a multi-stage optimization it will be important to capture the characteristic timescales of the processes that form the various stages. However, we note that calculating this measure is more computationally intensive than the asymptotic measure of efficiency we use in this work. This is because it requires temporal propagation of the system density matrix or population vector, whereas the asymptotic measure defined above only requires a rate matrix inversion. 

\end{document}